\documentclass[twocolumn]{aastex62}

\graphicspath{{./}{figures/}}

\shorttitle{CE Wind Tunnel: Applicability and Self-Similarity}
\shortauthors{Everson et al.}

\begin{document}

\title{Common Envelope Wind Tunnel: Range of Applicability and Self-Similarity in Realistic Stellar Envelopes}

\correspondingauthor{Rosa Wallace Everson}
\email{rosa@ucsc.edu}

\author[0000-0001-5256-3620]{Rosa Wallace Everson}
\altaffiliation{NSF Graduate Research Fellow}
\affiliation{Department of Astronomy \& Astrophysics, University of California, Santa Cruz, CA 95064, USA}
\affiliation{Niels Bohr Institute, University of Copenhagen, Blegdamsvej 17, 2100 Copenhagen, Denmark}

\author[0000-0002-1417-8024]{Morgan MacLeod}
\affiliation{Harvard-Smithsonian Center for Astrophysics, 60 Garden Street, Cambridge, MA 02138, USA}

\author[0000-0002-3316-5149]{Soumi De}
\affiliation{Department of Physics, Syracuse University, Syracuse, NY 13244, USA}
\affiliation{Kavli Institute for Theoretical Physics, University of California, Santa Barbara, CA 93106, USA}

\author[0000-0002-9946-4635]{Phillip Macias}
\affiliation{Department of Astronomy \& Astrophysics, University of California, Santa Cruz, CA 95064, USA}
\affiliation{Niels Bohr Institute, University of Copenhagen, Blegdamsvej 17, 2100 Copenhagen, Denmark}

\author[0000-0003-2558-3102]{Enrico Ramirez-Ruiz}
\affiliation{Department of Astronomy \& Astrophysics, University of California, Santa Cruz, CA 95064, USA}
\affiliation{Niels Bohr Institute, University of Copenhagen, Blegdamsvej 17, 2100 Copenhagen, Denmark}

\begin{abstract}
Common envelope evolution, the key orbital tightening phase of the traditional formation channel for close binaries, is a multistage process that presents many challenges to the establishment of a fully descriptive, predictive theoretical framework. In an approach complementary to global 3D hydrodynamical modeling, we explore the range of applicability for a simplified drag formalism that incorporates the results of local hydrodynamic ``wind tunnel" simulations into a semi-analytical framework in the treatment of the common envelope dynamical inspiral phase using a library of realistic giant branch stellar models across the low, intermediate, and high mass regimes. In terms of a small number of key dimensionless parameters, we characterize a wide range of common envelope events, revealing the broad range of applicability of the drag formalism as well its self-similar nature  across mass regimes and ages. Limitations arising from global binary properties and local structural quantities are discussed together with the opportunity for a general prescriptive application for this formalism.
\end{abstract}


\section{Introduction} \label{sec:intro}

It is well known that stars, rather than forming singly, are often formed in a binary or a triple system in which the stars orbit about their mutual center of mass \citep[e.g.][]{Sana2012, Toonen2016}. The evolution and fate of individual main sequence stars are well understood, and in multi-body systems in which the stars are separated by large distances relative to their sizes, we expect them to evolve much as they would alone. However, the evolution of binary systems in which the stars are close enough to interact is not as well understood, largely due to the countless variations of possible parameters: initial separation, mass ratio, evolutionary stage, and so forth. Though we may establish limits to these parameters via observation, such limits are constrained largely to local short-period systems (close binaries). However, close binaries in general are of great interest due to their role as possible precursors to many types of high-energy transients \citep[see, e.g.][]{Bethe1998, Lee2007, 2011ApJ...737...89D, Postnov2014}, including binary neutron star and binary black hole mergers detected by LIGO \citep[e.g.][]{Abbott2019}.

All close binary systems in which stellar remnants orbit at a separation smaller than the radii of their progenitor stars must have undergone some type of orbital transformation. In high stellar density regions, dynamical interactions may be a viable formation channel for close binaries \citep[see, e.g.][]{Samsing2018, Rodriguez2018}, and in binaries that initially form close to contact, chemically homogeneous evolution may forego the need for any tightening \citep[see, e.g.][]{Mandel2016}, but in other cases orbital tightening of a pre-existing binary must be accomplished by one or more phases of common envelope (CE) evolution. A CE phase occurs when one member in a binary, hereafter called the primary, moves off the main sequence and expands beyond its Roche lobe, engulfing the other typically lower mass member, or secondary, and creates a system in which the core of the primary interacts with the secondary within a shared envelope \citep[e.g.][]{Paczynski1976, Taam2000, Taam2010, Ivanova2013, Iben1993,2020arXiv200109829V}. Though the primary is always a star in its giant phase, the secondary may be a planet, a lower mass main sequence star, or any kind of stellar remnant. 

Though several stages of CE evolution may occur for a given binary, there are only two final outcomes: either the envelope is ejected and binarity is preserved, or the envelope is not fully ejected and the secondary merges with the core of the primary. The structure of the envelope and the properties of the embedded secondary both play a role in deciding the outcome of CE evolution; decades of analytical and computational study have provided insight into precisely how, but still leave many questions unanswered \citep[see][for extensive reviews]{Ivanova2013, Ivanova2016a}.

Extensive work has been done to produce global 3D simulations of CE evolution \citep[e.g.][]{Ricker2008, Ricker2012, Passy2012, Nandez2014, Nandez2016, Ivanova2016, Ohlmann2016a, Ohlmann2016b, Ohlmann2017, Staff2015, Staff2016, Iaconi2016, Chamandy2018, Chamandy2019a, Chamandy2019b, Prust2019,Wu2020}, but these efforts have faced many challenges, including (but not limited to) resolving adequately at all relevant physical scales, which span many orders of magnitude. An alternative and complementary approach has been developed by \citet{MacLeod2015a, MacLeod2015b, MacLeod2017b}, and greatly extended by \citet{De2019} in a companion paper, to explore the local CE behavior around an embedded compact object using a ``wind tunnel'' morphology. This morphology, rather than modeling the plunge of the secondary through the envelope globally, focuses on a region centered on the (fixed) secondary in the interior of the envelope and subjects it to a wind representing the passing envelope material, reducing the relevant scales within the simulation domain. This is achieved numerically by modeling the secondary as a fixed, accreting compact object that is subject to a supersonic wind with a density structure consistent with polytropic extended stellar envelopes. Key flow parameters are described by specific dimensionless quantities as described in Section \ref{sec:flow}. Due to the use of Cartesian geometry, the ``wind tunnel'' approximation is appropriate only for systems in which the extent of gravitational influence of the embedded object on the envelope material is much less than the extent of the envelope itself.

The broad range of masses and configurations of systems that undergo CE evolution tend to be investigated in separate regimes due to the differences in possible outcomes, structure, and key physics of the objects that comprise each system. However, the dynamical inspiral phase appears to be governed by just a few dimensionless parameters (see Section \ref{subsec:scales}) that can be calculated for any and all configurations for which the ``wind tunnel'' approximation is appropriate. Any self-similarity that exists in these parameters, regardless of the global characteristics of the binary, can be exploited via their connection to drag forces and accretion rates \citep{De2019, MacLeod2015a, MacLeod2015b, MacLeod2017b} to constrain and inform models of the dynamical inspiral phase and binary properties at the end of that phase.

In this work, we examine a range of realistic stellar models in terms of these parameters to determine the range of applicability for the formalism of \citet{MacLeod2017b} and, by extension, the mapping of the results from \citet{De2019} to envelope parameters for the calculation of inspiral trajectories. In Section \ref{sec:evolution}, we discuss the relevant aspects of late stage stellar evolution across mass regimes, noting key features that differentiate these regimes. In Section \ref{sec:flow}, we present the flow parameters and numerical results that together makeup the ``drag formalism" as established by \citet{MacLeod2015a, MacLeod2015b, MacLeod2017b} for which we seek to establish firm limits of applicability. In Section \ref{sec:mapping}, we map a broad range of CE events into the parameter space defined by the drag formalism, detailing how the properties of realistic stellar envelopes allow for general use. We address in detail the limitations and exceptions that define the range of applicability in Section \ref{sec:applicability}, including the validity of our results across additional model parameters and indications that the drag formalism naturally differentiates inspiral phases. In Section \ref{sec:discussion}, we discuss how these results may be combined with those from \citet{De2019} to further application of the drag formalism.

\section{Properties of Evolved Stars} \label{sec:evolution}

In CE events, the primary has evolved beyond the main sequence into the giant branch. All stars in the giant branch have some structural similarities, namely extended, diffuse envelopes and a small, dense core that is no longer centrally burning hydrogen. However, the specifics of a given giant's structure vary widely depending on the mass and age of the star, in turn varying the applicable physics pertaining to energy transport in the envelope, distinguishing core from envelope, and of course, success or failure of envelope ejection, among other things. In exploring the limits of the drag formalism, which depends upon a few key dimensionless parameters, we first endeavor to understand which similarities and differences in familiar structural terms are relevant to the dynamical inspiral phase of CE.

\begin{figure}[tbp]
	\figurenum{1}
	\epsscale{1.1}
	\plotone{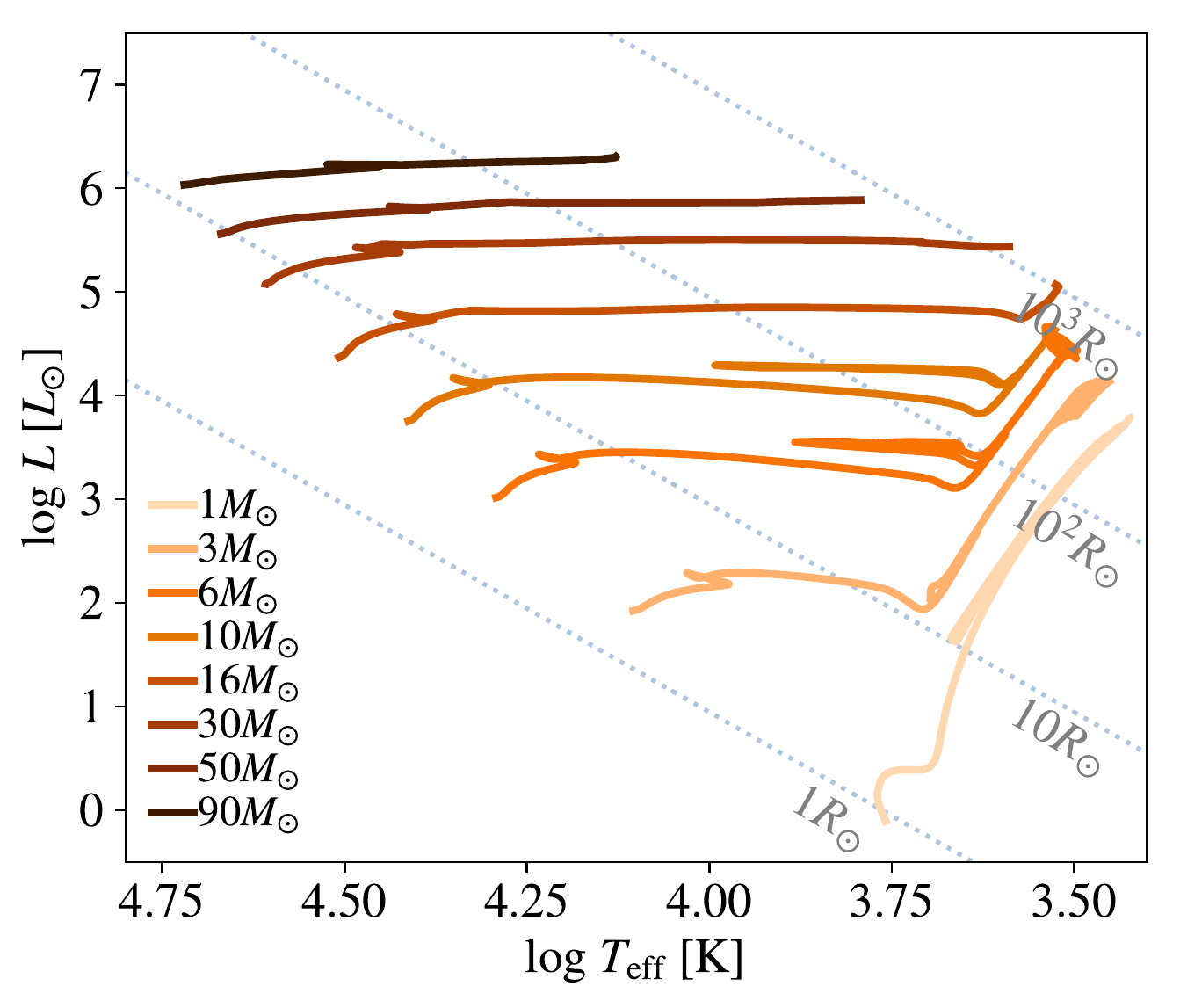}
	\caption{HR diagram of evolutionary tracks from ZAMS for a selection of primary stars used in this study. Contours of fixed radius are shown. For a given initial separation and mass ratio, primaries of vastly different mass are able to initiate a CE phase at some point in their post-main sequence evolution. However, the corresponding differences in envelope structure impact the dynamics and outcome of the CE phase in fundamental ways.} \label{fig:HRdiagram}
\end{figure}

The HR diagram shown in Figure \ref{fig:HRdiagram} traces the evolution from the zero-age main sequence (ZAMS), as simulated using the MIST package with MESA (for details, see Subsection \ref{subsec:methods}), of a selection of stars across a mass range that spans two orders of magnitude. Stars of vastly different mass and evolutionary track can expand to similar extent, with implications for the traditional formation channel of close binaries and CE evolution. Stars of different mass will reach the same extent at different stages of their giant branch, with corresponding differences in envelope structure related to mass and evolutionary stage.

\begin{figure}[tbp]
	\figurenum{2}
	\epsscale{1.1}
	\plotone{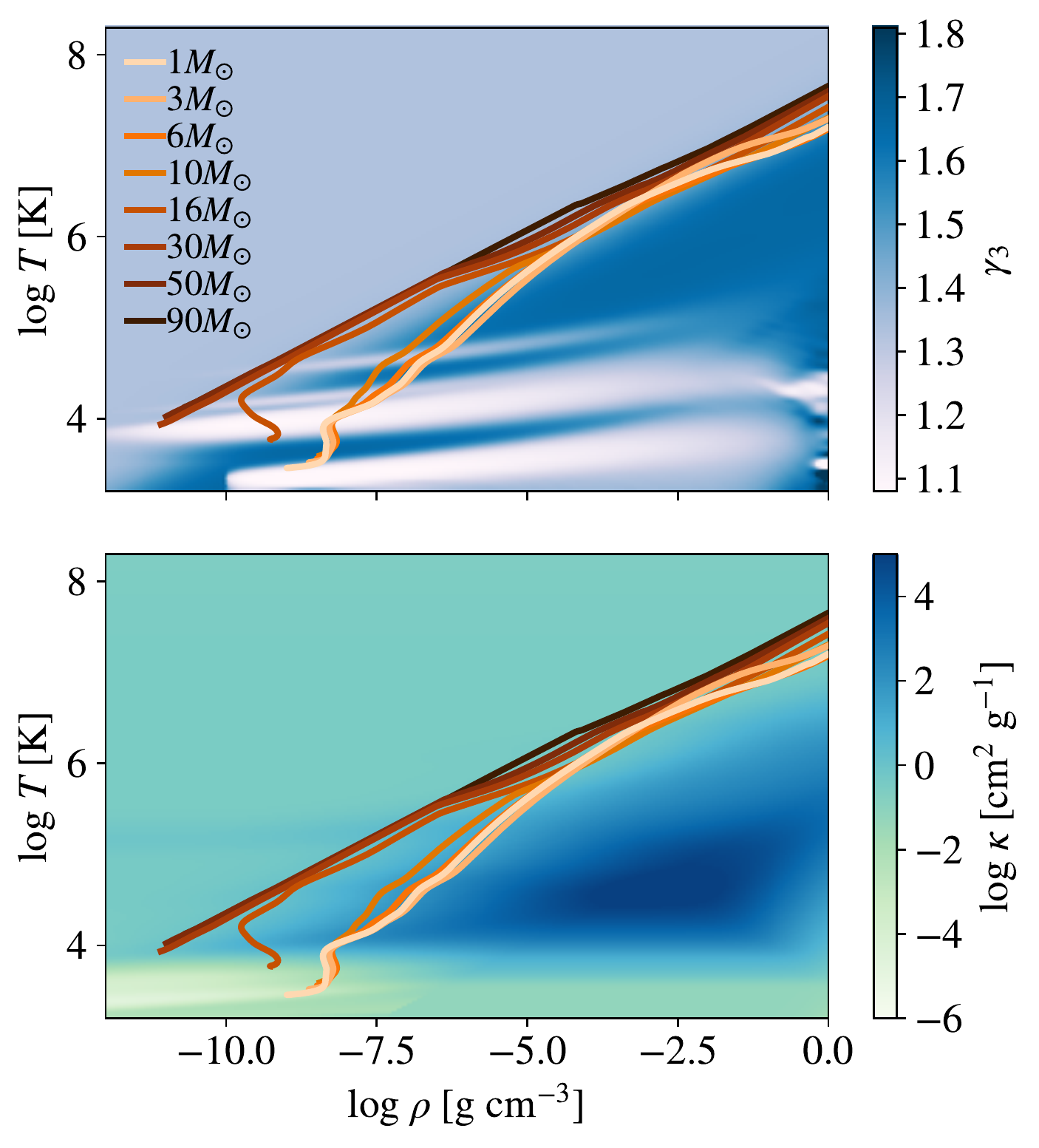}
	\caption{The selection of stars from Figure \ref{fig:HRdiagram} shown in the $\rho-T$ plane during the giant branch at $\approx 250 R_{\odot}$. Initial stellar masses are labeled. Adiabatic index is shown in blue in the upper panel and log opacity is shown in blue and green in the lower panel. The core of each star lies outside the upper right of each panel while envelopes are shown. Density inversions near the limb are seen in the stars with mass $\leq 16 M_\odot$ due to hydrogen and helium opacity peaks. In the upper panel, regions of low adiabatic index correspond to zones of partial ionization. For extended stars, the envelope equation of state tends to be dominated by convection ($\gamma \sim 5/3$) in lower mass stars and radiation pressure ($\gamma \sim 4/3$) in higher mass stars, seen here in the shift to lower adiabatic index for tracks of increasing mass. In the lower panel, the envelopes of more massive stars can be seen to have fairly constant opacity throughout, with more variability in those of lower mass stars.}  \label{fig:rhoTintro}
\end{figure}

To make such a comparison, we look at a range of stars that have all reached an extent of $\approx 250 R_{\odot}$. In Figure \ref{fig:rhoTintro}, envelopes are shown in the $\rho-T$ plane overplotted against adiabatic index and opacity values. Notably, none of the envelopes shown could be described as perfectly polytropic. In fact, the outer envelope often contains one or more regions of highly compressible material interspersed with convective or radiative regions, including density inversions that correspond to hydrogen and helium opacity peaks at $T \sim 5500 \mathrm{K}$ and $13000 \mathrm{K}$ \citep{Sanyal2015, Guzik2018}. The differences in structure seen here affect key processes in CE evolution, namely orbital decay due to drag and the ability of released energy to escape the envelope \citep[see, e.g.][]{Wilson2019, Wilson2020, Grichener2018}. How impactful these differences are on CE inspiral, however, is dependent on how much of the envelope contains these variations. 

\begin{figure}[tbp]
	\figurenum{3}
	\epsscale{1.1}
	\plotone{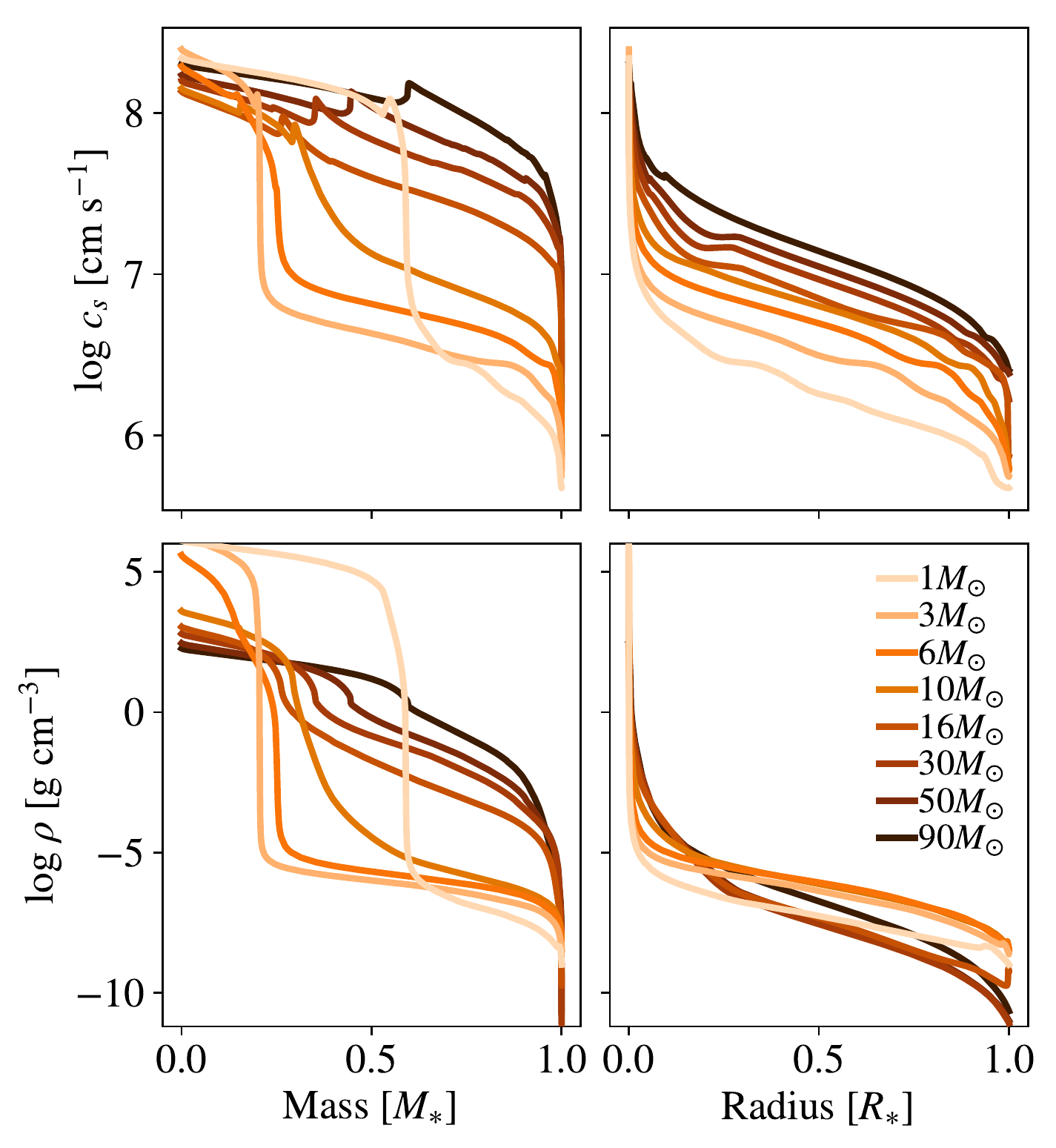}
	\caption{Selected giant branch stars extending to $\sim 250 R_{\odot}$, with upper panels showing sound speed $c_s$ and lower panels showing density in mass and radius coordinates. Differences in structure shown in mass coordinates are less apparent in radial coordinates, as these relate to core structure and how much relative mass is contained in the core and envelope respectively. Density structure through the envelopes of all the stars shown is similar, with sound speed increasing with mass due to a corresponding increase in luminosity/temperature, as seen in Figures \ref{fig:HRdiagram} \& \ref{fig:rhoTintro}.}  \label{fig:csrhointro}
\end{figure}

In Figure \ref{fig:csrhointro}, we examine the structure of the same stellar profiles seen in Figure \ref{fig:rhoTintro} in terms of the familiar structural quantities of sound speed and density against mass and radius. In mass coordinates, we can clearly see how the mass of each star is distributed differently, even amongst stars in the same mass regime. In the lower mass stars, the core-envelope boundary can be identified as a steep increase in density/sound speed, while in the higher mass stars a sharp, local peak in sound speed is the clearest indicator. This gives a sense of how much mass is held in the envelope, hence where the most binding energy lies within the star, and that the region relevant for CE inspiral contains only a fraction of the star's total mass, often less than half. In radial coordinates, the envelopes look similar in density structure, though the differences in sound speed impact the orbital decay during inspiral. Worth noting, however, are the minor density inversions that occur very close to the limb of most of these models, which coincide with the regions of highly compressible gas seen in white in Figure \ref{fig:rhoTintro} and are an important consideration when choosing how to apply the drag formalism (for details, see Subsection \ref{subsec:EOS}).

\section{Flow Parameters in the Stellar Envelope} \label{sec:flow}

The application of simulation results from \citet{De2019, MacLeod2015a, MacLeod2017b} requires that we interpret envelope structure as it relates to CE inspiral using the dimensionless quantities used in those studies. This allows us to characterize a dynamic process that involves many relevant physical quantities and variations with a few key parameters that combine information about the structure of the envelope, properties of the binary, and inspiral mechanics. For additional details beyond the brief introduction given here, the reader is referred to \citet{MacLeod2015a, MacLeod2015b}  and \citet{MacLeod2017b}.

\subsection{Relevant Scales and Parameters} \label{subsec:scales}

We model our typical CE system in simplest terms as a binary in which the primary, with mass $M_1$, is more massive and extended, and the secondary is a compact, lower mass object of mass $M_2$. We define the global mass ratio of the binary as $q_\mathrm{B} = M_2 / M_1$. The center of the primary is separated from the secondary by a distance $a$. At any given point after the onset of CE, the primary mass enclosed at separation $a$ is defined as $M_\mathrm{enc}<M_1$. We define the mass ratio between the secondary and the mass enclosed at separation $a$ as
\begin{equation}
    q_\mathrm{r} = \frac{M_2}{M_\mathrm{enc}}.
\end{equation}
This quantity will increase as inspiral progresses since  $M_\mathrm{enc}$ decreases with $a$, though this is most pronounced in the inner envelope. Any accretion onto the secondary will serve to boost this effect. 

Following the formalism of \citet{MacLeod2017b}, we approximate our inspiral to first order as a modified keplerian orbit, giving the velocity of the secondary relative to the envelope material as
\begin{equation}
    v_{\infty} = f_\mathrm{k} \sqrt{\frac{G (M_\mathrm{enc} + M_2)}{a}}
\end{equation}
in which $f_\mathrm{k}$ reflects the degree to which the rotation of the envelope and the orbit of the secondary are non-synchronous (ie. $f_\mathrm{k} =1$ gives a perfectly keplerian orbit with no co-rotation, and $f_\mathrm{k} =0$ gives an orbit in which the envelope and secondary are tidally locked). 

Moving into dimensionless terms, we use the framework for flows and accretion first introduced by \citet{Hoyle1939}, hereafter HLA \citep{Hoyle1939,Bondi1944}. We characterize the relative velocity $v_\infty$ with Mach number
\begin{equation}
    \mathcal{M}_\infty = \frac{v_{\infty}}{c_\mathrm{s}}
\end{equation}
in which $c_s$ is the local sound speed of the undisturbed envelope material at separation $a$. Generally, dynamical inspiral spans a range of low Mach numbers, on order of a few. As the secondary moves through the envelope, it will affect oncoming material gravitationally as it passes by; the cross-section of oncoming material that is within this gravitational ``sphere of influence'' is characterized by the accretion radius
\begin{equation}
    R_\mathrm{a} = \frac{2 G M_2}{v_{\infty}^2}
\end{equation}
which is a function of not only the secondary's mass, but the changing enclosed mass and separation $a$. To get a sense of how strong the impact of an envelope density gradient may be on the flow and accretion, we compare $R_\mathrm{a}$ to the density scale height at the location of the secondary
\begin{equation}
    H_{\rho} = -\rho \frac{dr}{d\rho} ,
\end{equation}
which describes the local density normalized by the local density gradient with respect to radius. From this comparison arises the quantity
\begin{equation}
    \epsilon_{\rho} = \frac{R_\mathrm{a}}{H_{\rho}} ,
\end{equation}
which is a measure of how many local scale heights are traversed by the local accretion radius (ie. $\epsilon_{\rho} = 0$ corresponds to a constant density medium and a symmetric HLA-type flow and accretion, while $\epsilon_{\rho} > 1$ corresponds to density gradients that break the symmetry in the flow and suppress accretion significantly).

\subsection{Envelope Equation of State} \label{subsec:envelopeEOS}

The drag formalism was developed with the assumption of a polytropic envelope, out of which arises a structural polytropic index,
\begin{equation}
    \Gamma_\mathrm{s} = \bigg(\frac{d \ln P}{d \ln \rho}\bigg)_\mathrm{env},
\end{equation}
which is evaluated along the envelope profile, such that $P \propto \rho^{\Gamma_\mathrm{s}}$. For MIST/MESA stellar profiles, we smooth the numerical derivative with a Gaussian filter with standard deviation of $\sim 1 \%$ of the envelope's extent.

However, the envelope material does not always behave as an ideal gas, requiring an equation of state (EOS) with several adiabatic indices to describe its behavior when compressed during the inspiral of the secondary. The indices of interest to us are as follows:
\begin{equation}
    \gamma_1 = \bigg(\frac{d \ln P}{d \ln \rho}\bigg)_\mathrm{ad},
\end{equation}
which is used to compute the local sound speed, and
\begin{equation}
    \gamma_3 - 1= \bigg(\frac{d \ln T}{d \ln \rho}\bigg)_\mathrm{ad},
\end{equation}
which is used to relate pressure, density, and internal energy. These indices are the same in an ideal gas, and are equivalent to $\Gamma_\mathrm{s}$ at constant entropy.  

\begin{figure*}[tbp]
	\figurenum{4}
	\epsscale{1.1}
	\plotone{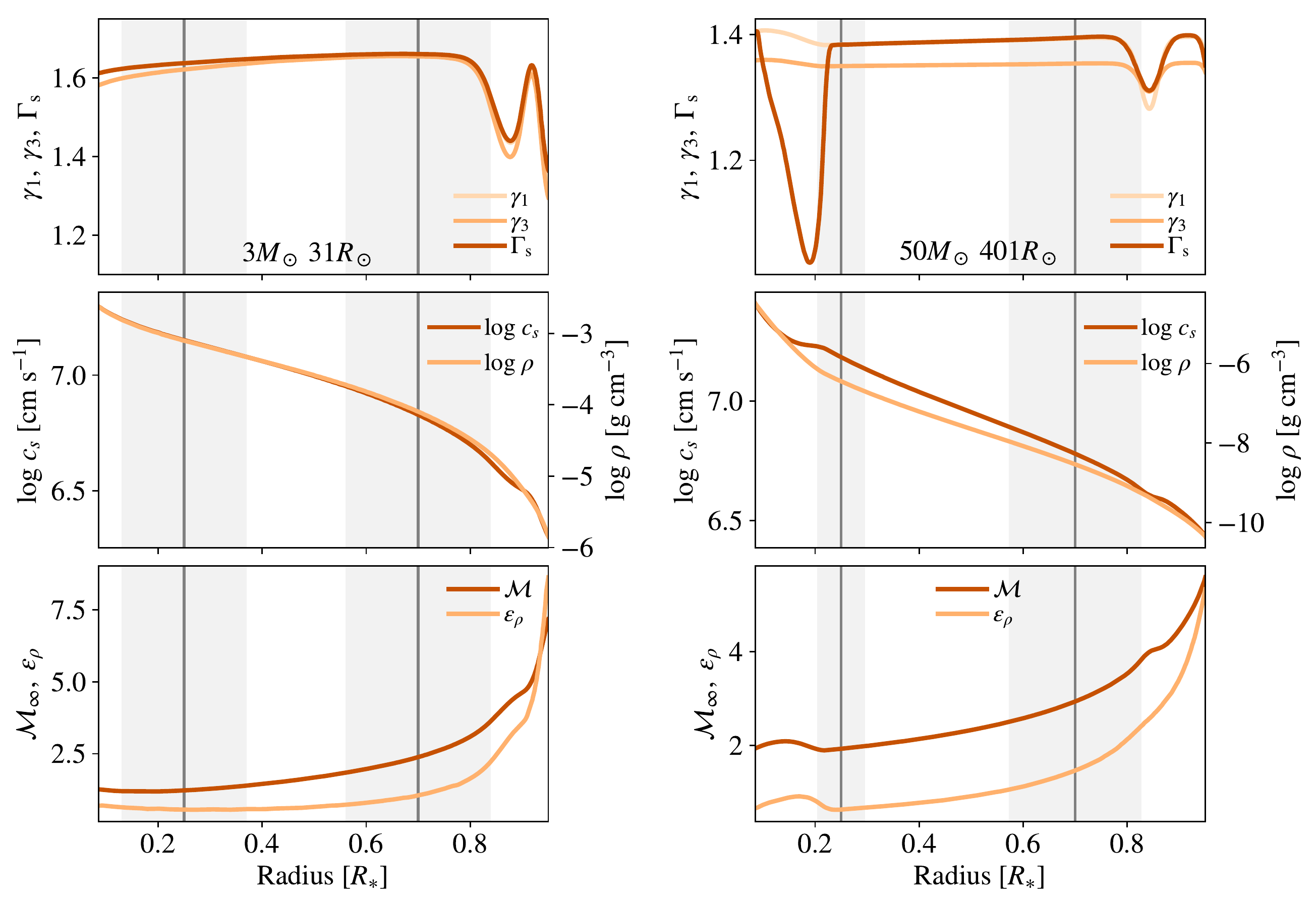}
	\caption{A comparison of standard EOS and structural quantities against drag formalism parameters for initial mass $3 M_\odot$ and $50 M_\odot$ stars at $31 R_\odot$ and $401 R_\odot$, respectively, with a secondary of mass ratio $0.1$. Only the envelope is shown. For the secondary located at a given dark grey line, shaded regions show the span of $R_\mathrm{a}$ to the left and right of that location. Note that $R_\mathrm{a}$ has a location dependence. The extent of this region illustrates the envelope material that is gravitationally influenced by the secondary during inspiral. \textit{Left panels:} The envelope in this case is largely convective, the majority having $\Gamma_\mathrm{s} \sim 5/3$ with some higher compressibility regions in the outer part. In convective envelopes, we expect $\Gamma_\mathrm{s} \sim \gamma_1 \sim \gamma_3$ due to constant entropy. \textit{Right panels:} The envelope is largely radiative, giving different values for $\gamma_1$ and $\gamma_3$: $\Gamma_\mathrm{s} \sim \gamma_1 \sim 1.4$ until approaching the core, but $\gamma_3$ maintains a slightly lower value $\sim 4/3$.}  \label{fig:comparison}
\end{figure*}

In Figure \ref{fig:comparison}, we present for comparison these three indices, as well as the familiar structural quantities of sound speed $c_s$ and density $\rho$, with the corresponding $\mathcal{M}_\infty$ and $\epsilon_\rho$ values calculated for MIST models of initial mass 3 and $50 M_\odot$, respectively, with a secondary of mass ratio $q_\mathrm{B}=0.1$. Throughout most of the envelope in both cases, $\Gamma_\mathrm{s} \sim \gamma_1$, with noted exception upon approaching the core. Features are naturally mirrored among all of these quantities, to a greater or lesser extent, yet the monotonic decrease we would expect in $\mathcal{M}_\infty$ and $\epsilon_\rho$ from the limb to the core for a polytropic envelope is still represented here. Therefore we cautiously move forward with a simplified approach that may allow us to parameterize dynamical inspiral further.

\subsection{Polytropic Formalism} \label{subsec:formalism}

When assuming a polytropic stellar profile, the relationships of the flow parameters of Subsections \ref{subsec:scales} and \ref{subsec:envelopeEOS} can be constructed in the following manner, as in \citet{MacLeod2017b}:

\begin{equation}
    \epsilon_{\rho} = \frac{2 q_\mathrm{r}}{(1+q_\mathrm{r})^2} \frac{\mathcal{M}_\infty^2}{f_\mathrm{k}^4} \bigg(\frac{\gamma_1}{\Gamma_\mathrm{s}}\bigg).
\end{equation}

In the simplified case in which the inspiral velocity is approximated to first order as keplerian and the envelope has constant entropy, this expression simplifies to
\begin{equation}\label{eq:polyrel}
    \epsilon_\rho = \frac{2q_\mathrm{r}}{(1+q_\mathrm{r})^2} \mathcal{M}_\infty^2 .
\end{equation}
This implies that for these special cases, our flow parameters are intrinsically linked, and that two of these quantities may be sufficient to characterize the flow at a given location.

\subsection{Key Results of Hydrodynamic Simulations} \label{subsec:simresults}

Using a traditional HLA framework, the drag force on the secondary is expected to be
\begin{equation}
    F_\mathrm{d,HLA} = \pi R_\mathrm{a}^2 \rho_\infty v_\infty^2 
\end{equation}
and the corresponding accretion rate on to the secondary
\begin{equation}
    \dot{M}_\mathrm{HLA} = \pi R_\mathrm{a}^2 \rho_\infty v_\infty,
\end{equation}
in which $\rho_\infty$ is the density of undisturbed oncoming wind. However, these expressions assume $\rho_\infty$ to be constant, and are unlikely to match that measured when a wind with a density gradient is used and symmetry in the wake is broken. A key result from the suite of simulations performed by \citet{MacLeod2015a, MacLeod2017b, De2019} is a grid of drag force measurements 
\begin{equation}
    F_\mathrm{d} = C_\mathrm{d} F_\mathrm{d,HLA}
\end{equation} 
and accretion rates
\begin{equation}
    \dot{M} = C_\mathrm{a} \dot{M}_\mathrm{HLA}
\end{equation}
in which $C_\mathrm{d}$ and $C_\mathrm{a}$ are drag and accretion coefficients, respectively, that characterize the steady-state time-averaged drag force and accretion rate from a specific simulation setup normalized by the calculated HLA values based on the undisturbed envelope density $\rho$ at the location of the secondary. As each simulation setup reflects a single value for each of $\mathcal{M}_\infty$, $q_\mathrm{r}$,  $\epsilon_\rho$, and $\gamma$ (for setups in which $\gamma=\Gamma_\mathrm{s}=\gamma_1=\gamma_3$), each pair of $C_\mathrm{d}$ and $C_\mathrm{a}$ then maps to a specific combination of these four quantities, all of which may be calculated or approximated with a basic stellar model and global properties of the pre-CE system.

These coefficients form the basis for broad application of the drag formalism to any type of CE event that may be of interest, using the above quantities to map coefficient values via interpolation or fitting functions. Examples include integration of the equation of motion of dynamical inspiral using a static stellar model \citep[e.g. Figures 11, 12 of][]{MacLeod2017b}, introduction of a heating term in 1D hydrodynamic simulations of CE \citep{Fragos2019} through the relation $\dot{E} \approx F_\mathrm{d} v_\infty$ \citep{MacLeod2015a}, and calculation of drag force for comparison against that produced by global 3D hydrodynamic simulations \citep{Chamandy2019a}. Notably, \citet{Chamandy2019a} found that during dynamical inspiral, when the assumptions of the drag formalism are met, the drag force calculated with the coefficients is in excellent agreement with that measured in a global simulation. This encouraging result shows the drag formalism to be an effective prescription for dynamical inspiral, and motivates its further development.

\section{Mapping of Dynamical Inspiral in Simulation Parameter Space} \label{sec:mapping}

Any binary system that results in a merger or close binary via the traditional formation channel must go through at least one CE phase. The flow parameters discussed in Section \ref{sec:flow} allow us to analyze CE inspiral not in terms of familiar quantities (ie. $\rho$, $v_\mathrm{orb}$, etc.) that keep structural and dynamical information separate, but in terms of dimensionless quantities (ie. $\epsilon_\rho$, $\mathcal{M}_\infty$, $q_\mathrm{r}$) that combine properties of the system with local structure and dynamics. Translated into the latter, a given system's inspiral corresponds to a curve in parameter space that traces the evolution of the three flow parameters from the outer regions of the envelope to a transitional region near the core boundary. Each point in this parameter space corresponds to a unique drag coefficient $C_\mathrm{d}$ and accretion coefficient $C_\mathrm{a}$ (see Subsection \ref{subsec:simresults}) that, when included in inspiral calculations, alters the orbital decay expected from the HLA formalism. By understanding the curves through this parameter space for a range of different progenitor systems, we can apply these drag coefficients to any stellar envelope based on the properties derived directly from stellar models.

\subsection{Methodology}
\label{subsec:methods}

In utilizing the results of the numerical simulations from \citet{MacLeod2015a}, \citet{MacLeod2017b}, and \citet{De2019}, we assume progenitor systems that span a wide range of mass, age, internal structure, and separation that include one giant branch star (hereafter, the primary) and one compact  star (hereafter, the secondary) with $q_\mathrm{B}$ ranging from $0.1-0.35$. To ensure that the envelope material encountered is structured consistently with the simulations, we limit the range of mass ratio such that throughout dynamical inspiral the accretion radius of the secondary does not exceed the remaining separation. We generate a library of stellar models spanning the aforementioned axes using the MESA Isochrones and Stellar Tracks (MIST) \citep{Dotter2016, Choi2016} package with MESA v7503 \citep{Paxton2011, Paxton2013, Paxton2015}. We have chosen MIST models for analysis due to the observational calibrations of the framework, though there are limitations to its use at very high mass and low metallicity (see Subsection \ref{subsec:z}, in which we address results from alternative libraries).

To get an agnostic view of flow parameters across a range of binary systems, we evolve stars of solar metallicity from $1-90 M_\odot$ through the giant branch, including profiles in our analysis based on the criterion of increasing radius in log space (a proxy for binary separation at onset) up to the maximum radius ($R_\mathrm{max}$) produced by the code. Due to mass dependent differences in late stage evolution as well as winds/mass loss, the maximum radius of each primary is unique; for any system, the maximum possible separation for which a CE phase will occur is defined to first order by this value. 

CE inspiral takes place only in the stellar envelope, therefore we limit our analysis to that region. Dynamical inspiral occurs after CE onset, which disrupts the outer layers of the envelope \citep{MacLeod2017a}. Due to the ``wind tunnel" morphology that the drag formalism is based upon, in particular the presence and undisturbed structure of oncoming material, it is appropriate for use only after the secondary is embedded. Thus we begin our analysis at a very conservative limit of $a=0.95 R_\ast$ for each model as an ersatz starting point for the dynamical plunge of the secondary, which is considered to be embedded and desynchronized post-onset (see Subsection \ref{subsec:onset}), and stop our analysis outside the core (see Subsection \ref{subsec:hook}). Due to the uncertainties regarding the conditions for successful envelope ejection, we make no claims about the termination of our calculated inspirals in connection with the outcome of a given CE event. Rather, we choose to map the entire range in which the drag formalism might be applied, and discern general trends as well as the region of parameter space in which the formalism breaks down. 

Combining the global properties and structural quantities from our realistic stellar models with a range of $q_\mathrm{B}$ values with constant (non-accreting) $M_2$, we then calculate the drag formalism parameters to produce characteristic curves for each inspiral in the parameter space.

\subsection{Characteristic Curves of Dynamical Inspiral}
\label{subsec:curves}

\begin{figure*}[tbp]
    \figurenum{5}
    \epsscale{1.1}
    \plotone{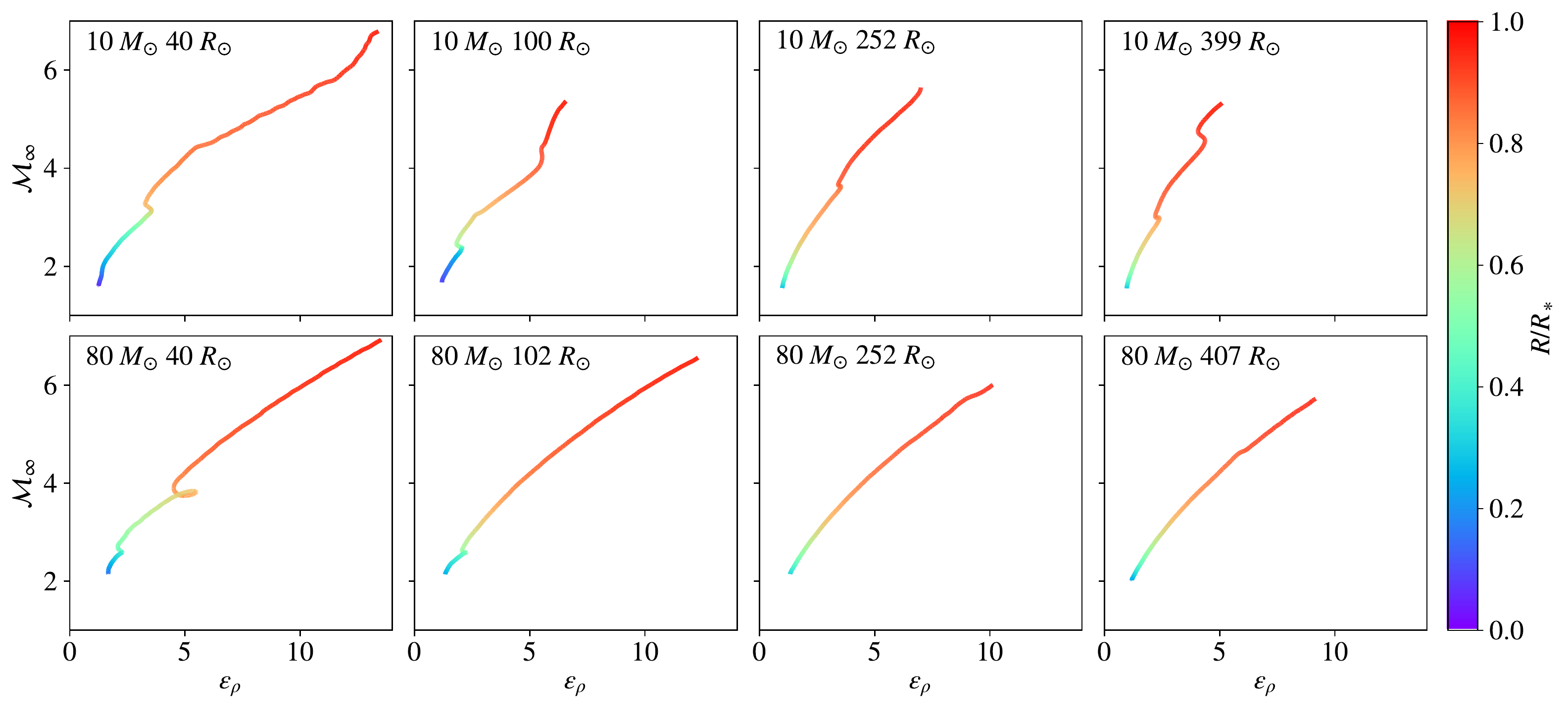}
    \caption{Representative examples of characteristic curves for CE events with $q_\mathrm{B} = 0.2$ in the $\mathcal{M}_\infty - \epsilon_\rho$ parameter space. In panels from left to right, stellar models increase in age and extent, with models of initial mass $10 M_\odot$ and $80 M_\odot$ represented on the top and bottom rows respectively. The upper right portion of each curve represents conditions in the outer envelope and the lower left portion of each curve represents conditions in the inner envelope, with normalized radius mapped in color. Each point on a curve corresponds to unique drag and accretion coefficients, making each characteristic curve a mapping of the dynamics occurring during a dynamical inspiral phase consistent with the setup, ie. primary and secondary masses, separation at onset, etc. This curve can be calculated for any appropriate binary with a sufficiently detailed stellar model for the primary.} \label{fig:exampletracks_rstar}
\end{figure*}

The shape of the characteristic curve for a given dynamical inspiral in the $\mathcal{M}_\infty - \epsilon_\rho$ parameter space is influenced by the structure of the envelope of the primary. In Figure \ref{fig:exampletracks_rstar}, we show selected curves for events with mass ratio $q_\mathrm{B}=0.2$ from various stages in the time evolution of initial mass $10 M_\odot$ and $80 M_\odot$ giants for comparison. These correspond to a range of binary separations: each panel represents a CE inspiral initiating at a separation equal to the model's extent, noted at the top of each panel. The color of the curve reflects the region in the extended primary where each set of combined $(\mathcal{M}_\infty, \epsilon_\rho)$ conditions exist in radial coordinates, with inspiral proceeding from the upper right to lower left corner of the parameter space. 

In general, inspiral is characterized by the highest values and broadest ranges of $\mathcal{M}_\infty$ and $\epsilon_\rho$ in the outer envelope, with lower values in the inner half of the envelope by radius. Though each curve is distinct, features which are present due to fluctuations in the envelope EOS (see Subsection \ref{subsec:EOS}) are minor.

\begin{figure}[tbp]
    \figurenum{6}
    \epsscale{1.15}
    \plotone{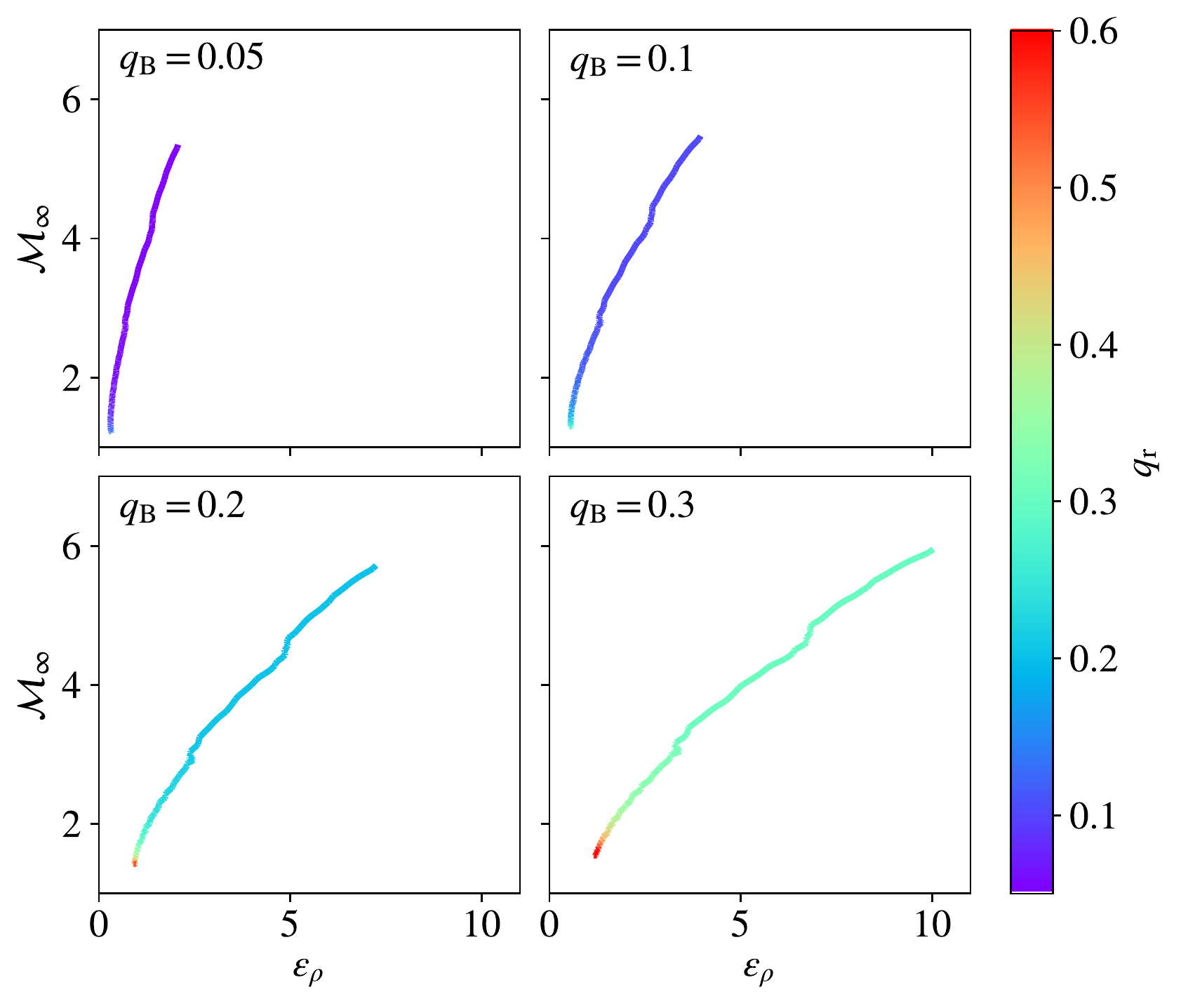}
    \caption{Additional examples of characteristic curves for CE events in the $\mathcal{M}_\infty - \epsilon_\rho$ parameter space involving a primary of $6 M_\odot$ at $250 R_\odot$ and a selection of $q_\mathrm{B}$ values. Curve color corresponds to the local value of $q_\mathrm{r}$. Due to the diffuse nature of envelope material, $q_\mathrm{r}$ is nearly constant until the secondary approaches the core. The slope of the curve decreases with increasing $q_\mathrm{B}$, reaching slightly higher $\mathcal{M}_\infty$ values and notably higher $\epsilon_\rho$ values for the same primary. This is due to the effect of the increase in $M_2$ on orbital velocity $v_\infty$ and the accretion radius $R_\mathrm{a}$, respectively.} \label{fig:exampletracks_qr}
\end{figure}

The effect of mass ratio on inspiral characteristic curves is shown in Figure \ref{fig:exampletracks_qr}. Using an example primary of $6 M_\odot$ evolved to $250 R_\odot$, we calculate curves for mass ratios $q_\mathrm{B}=0.05$, $0.1$, $0.2$, and $0.3$, which might represent, for example, a white dwarf, neutron star, or companion main sequence star secondary. There is a clear inverse relation between the slope of the curve and $q_\mathrm{B}$ value. According to the drag formalism, each point on a curve corresponds to a $C_\mathrm{d}$ and $C_\mathrm{a}$ value; however, these coefficients depend on the local mass ratio $q_\mathrm{r}$ to be correctly applied. Figure \ref{fig:exampletracks_qr} demonstrates that $q_\mathrm{r}$ remains nearly constant for the duration of dynamical inspiral, increasing appreciably only when the secondary reaches the innermost regions of the envelope. Thus we may justify a simplified application using something like an average mass ratio as by \citet{De2019}, especially when energy considerations indicate an outcome of successful envelope ejection, therefore avoiding the material near the core.

\subsection{Self-Similarity Across Axes} \label{subsec:selfsim}

\begin{figure*}[tbp]
	\figurenum{7}
	\epsscale{1.1}
	\plotone{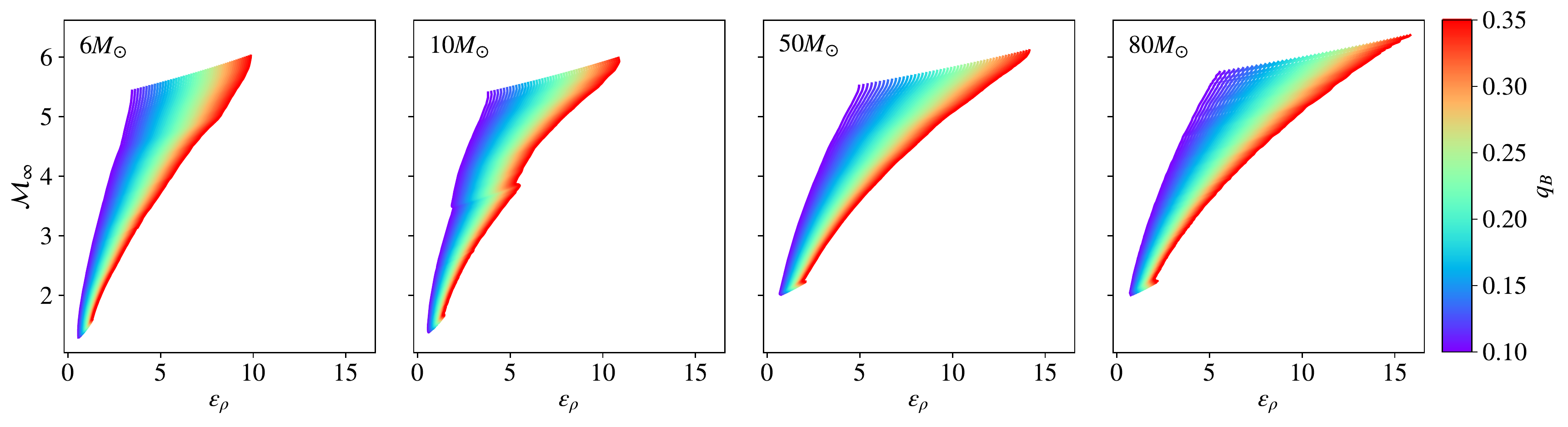}
	\caption{Characteristic curves of dynamical inspiral for primaries of initial mass 6, 10, 50, and 80$M_\odot$ and binary separation/radial extent of $250 R_\odot$. An appropriate range of $q_\mathrm{B}$ values for application of the drag formalism are plotted by color. Curves are shown to be self-similar in the $\mathcal{M}_\infty - \epsilon_\rho$ parameter space.} \label{fig:fanpanel}
\end{figure*}

In Figure \ref{fig:fanpanel}, we produce characteristic curves for inspirals with a range of $q_\mathrm{B}$ appropriate for the drag formalism across the axis of mass. The primary profiles used are giant stars of initial mass 6, 10, 50, and $80 M_\odot$ extended to $250 R_\odot$. These curves are representative across the entire library of stellar profiles, and repeat the trends seen in Figures \ref{fig:exampletracks_rstar} and \ref{fig:exampletracks_qr}. The decrease in slope corresponding to the increase in mass ratio combined with the similarity of these curves repeats this familiar fan shape throughout, and lends itself to further simplification.

 \begin{figure*}[tbp]
	\figurenum{8}
	\epsscale{1.1}
	\plotone{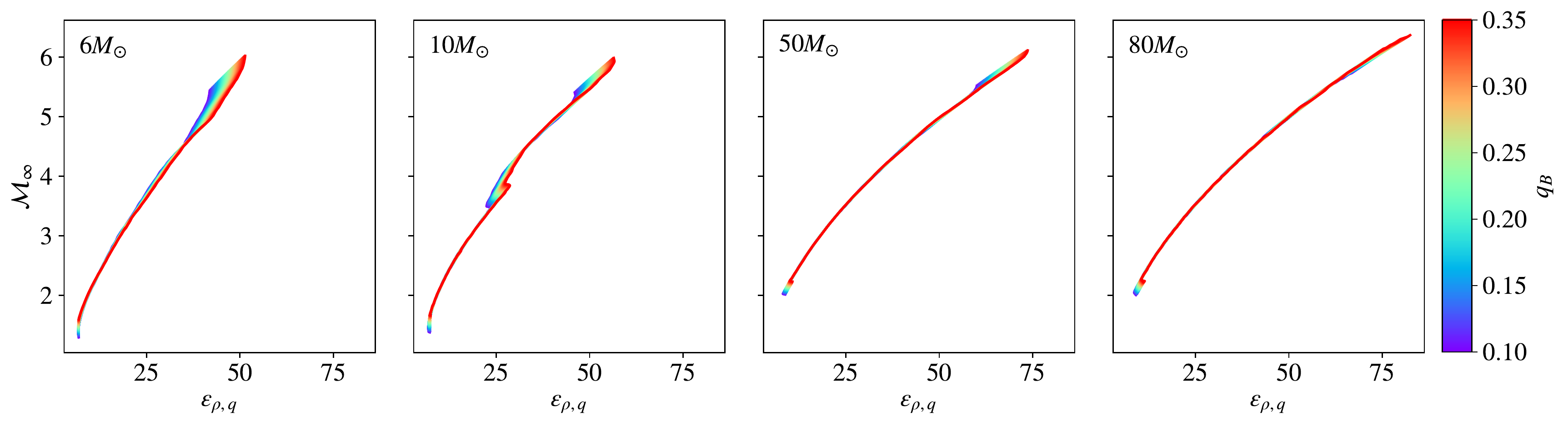}
	\caption{Shown are the same characteristic curves from Figure \ref{fig:fanpanel} normalized for the mass ratio term in Equations 10 and 11. In the ``collapsed'' $\mathcal{M}_\infty - \epsilon_{\rho,q}$ parameter space, dynamical inspirals for a given primary and separation are characterized by a single, nearly quadratic curve.} \label{fig:qnormfan}
\end{figure*}
\begin{figure*}[tbp]
    \figurenum{9}
    \epsscale{1.1}
    \plotone{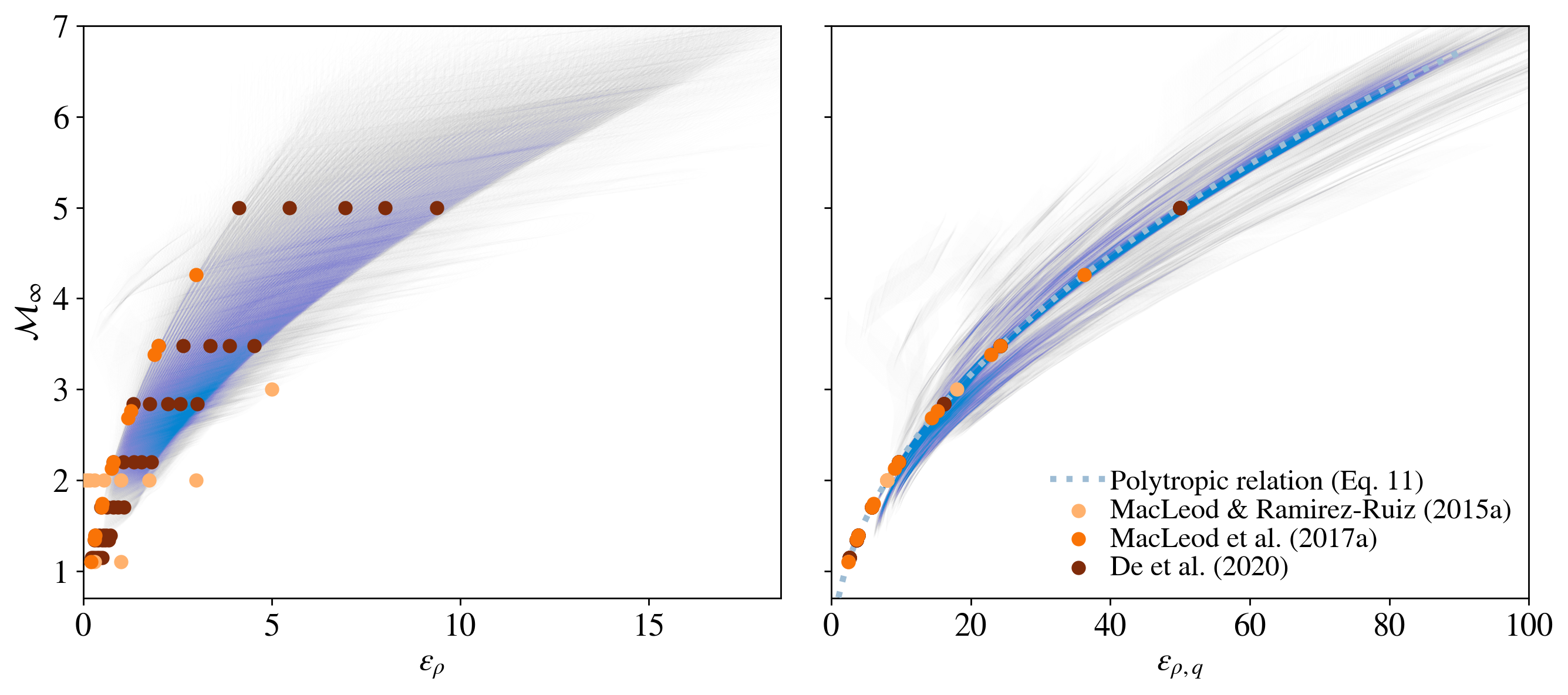}
    \caption{In the left panel are shown the overlaid characteristic curves for dynamical inspirals in the $\mathcal{M}_\infty - \epsilon_\rho$ parameter space for primaries of initial mass 1, 3, 6, 10, 16, 20, 30, 40, 50, 60, 70, 80, and 90 $M_\odot$ from the end of core H-burning, through each profile of increasing radius up to the maximum reached, for a range of $q_\mathrm{B}$ from 0.1-0.35. Transparent, grey regions in this parameter space have values less commonly encountered, while opaque, blue regions cover values that are extremely common. Overplotted are points in the parameter space for which local envelope drag and accretion coefficients have been calculated from ``wind tunnel'' simulations by \citet{MacLeod2015a, MacLeod2017b, De2019}. In the right panel are shown the same curves normalized over the mass ratio term, defined as $\epsilon_{\rho,q}$ in Equation 16. Consistency with the simplified polytropic relation of Equation 11, based on \citet{MacLeod2017b}, suggests an effective functional form requiring only a few envelope parameters may be possible to characterize the dynamical inspiral phase.}
    \label{fig:paramspace}
\end{figure*}

Using Equations 10 and 11 as our guide, we normalize these curves over the $q_\mathrm{r}$ term in Figure \ref{fig:qnormfan} using the following definition:
\begin{equation}
    \epsilon_{\rho,q} = \epsilon_\rho \frac{(1+q_\mathrm{r})^2}{q_\mathrm{r}}.
\end{equation}
Upon normalization, the fans collapse into a simple, approximately quadratic curve. Again, these curves are representative of the same calculations across the entire library of stellar profiles. The truncation of these curves in the outer envelope lie near $\mathcal{M}_\infty \sim 6$, and in the inner envelope are a function of how distinct the transition is from envelope to core, ranging from $\mathcal{M}_\infty \sim 1-2$.

In Figure \ref{fig:paramspace}, we repeat the above calculations for all post-main sequence stellar profiles from 1-90 $M_\odot$. The left panel reveals the region of the $\mathcal{M}_\infty - \epsilon_\rho$ parameter space that is represented in realistic stellar profiles and therefore ideal for simulation in order to support inspiral calculations more broadly. Though initial simulations by \citet{MacLeod2015a} and \citet{MacLeod2017b} cover the low mass ratio regime of this region, the full relevant parameter space is well-covered by the simulations in a companion paper, \citet{De2019}. Furthermore, in the right panel, the region in ``collapsed" parameter space that is most densely covered reveals the most basic characteristic curve for dynamical inspiral, and fits the polytropic relation of Equation~\ref{eq:polyrel}, which in such a broad range of non-polytropic envelopes reveals they are nonetheless polytropic ``enough'' for the drag formalism to be a good approximation of the conditions, and that there are fairly distinct truncation points to the overlay that we may take advantage of, to first order, in a prescriptive capacity.

Through a systematic comparison and analysis of characteristic curves across multiple axes, we find that nearly all dynamical inspirals that meet the basic criteria for application of the drag formalism (ie. $q_\mathrm{B} \lesssim 1/3$) are self-similar in $\mathcal{M}_\infty - \epsilon_\rho$ space. This self-similarity holds across the axes of primary mass $M_1$, the initial binary separation $a$ (or likewise the post-main sequence age/radius of the primary), and binary mass ratio $q_\mathrm{B}$ (for a discussion of the same across metallicity, see Subsection \ref{subsec:z}).

\section{Range of Applicability: Limitations and Exceptions}
\label{sec:applicability}

\subsection{Onset and Initial Mass Loss} \label{subsec:onset}

The dynamics of CE onset is an area of active study \citep[see, e.g.][]{Iaconi2016, MacLeod2018, Reichardt2019, Shiber2019, MacLeod2020a, MacLeod2020b} which is not yet well understood and has not yet been incorporated into the drag formalism. CE events occur after some initial destabilization of the binary: for some systems, this is a result of the Darwin tidal instability, and for others a result of unstable Roche lobe overflow \citep{MacLeod2017a}. The dependence on both mass ratio and primary stellar structure requires that both types of systems are represented in the range of binaries used in this work. The setup of the ``wind tunnel'' simulations assumes a plunge into undisturbed stellar envelope, and the envelope depth at which we may assume this criterion to be satisfied post-onset is variable and dependent on many factors which have  not been accounted for in a general formalism. 

Therefore, we choose to map the broadest range of envelope parameters, which assumes little or no mass loss prior to CE as in the case of Darwin instability, rather than removing large portions of envelope based on incomplete understanding. Incorporation of mass loss during onset, assuming no changes to the structure of the remaining envelope material, will bring the upper right truncation point of a given dynamical inspiral into a lower range of $\mathcal{M}_\infty$ and $\epsilon_\rho$ values, reducing the coverage of parameter space traced by that inspiral.

Depending on the duration of the pre-CE phase, the bound envelope material may adjust its structure relative to the static models used in this study. The application of this framework to such adjusted models would not have an impact on the drag and accretion coefficients as they correlate to the parameter space, but would simply change the extent and region of parameter space crossed during a particular dynamical inspiral relative to a static model. Due to the representation of a broad range of envelope configurations and their consistent tracing of the same parameter space, it is unlikely that these changes would push a characteristic curve outside of the region represented here. Further work is needed to explore the junction of onset mechanics and the drag formalism for self-consistent application.

\subsection{The Dynamical Boundary} \label{subsec:hook}

The appropriate definition of the core boundary for purposes of CE calculations is difficult to pinpoint for stars in different mass regimes and various stages of post-main sequence evolution.  In attempting to account for the varying criteria used in the literature to define that boundary \citep{Tauris2001, Ivanova2013}, we applied different definitions to characteristic curve calculations across the full model library and found that the drag formalism presents its own unique termination point - the dynamical boundary.

\begin{figure}[tbp]
    \figurenum{10}
    \epsscale{1.1}
    \plotone{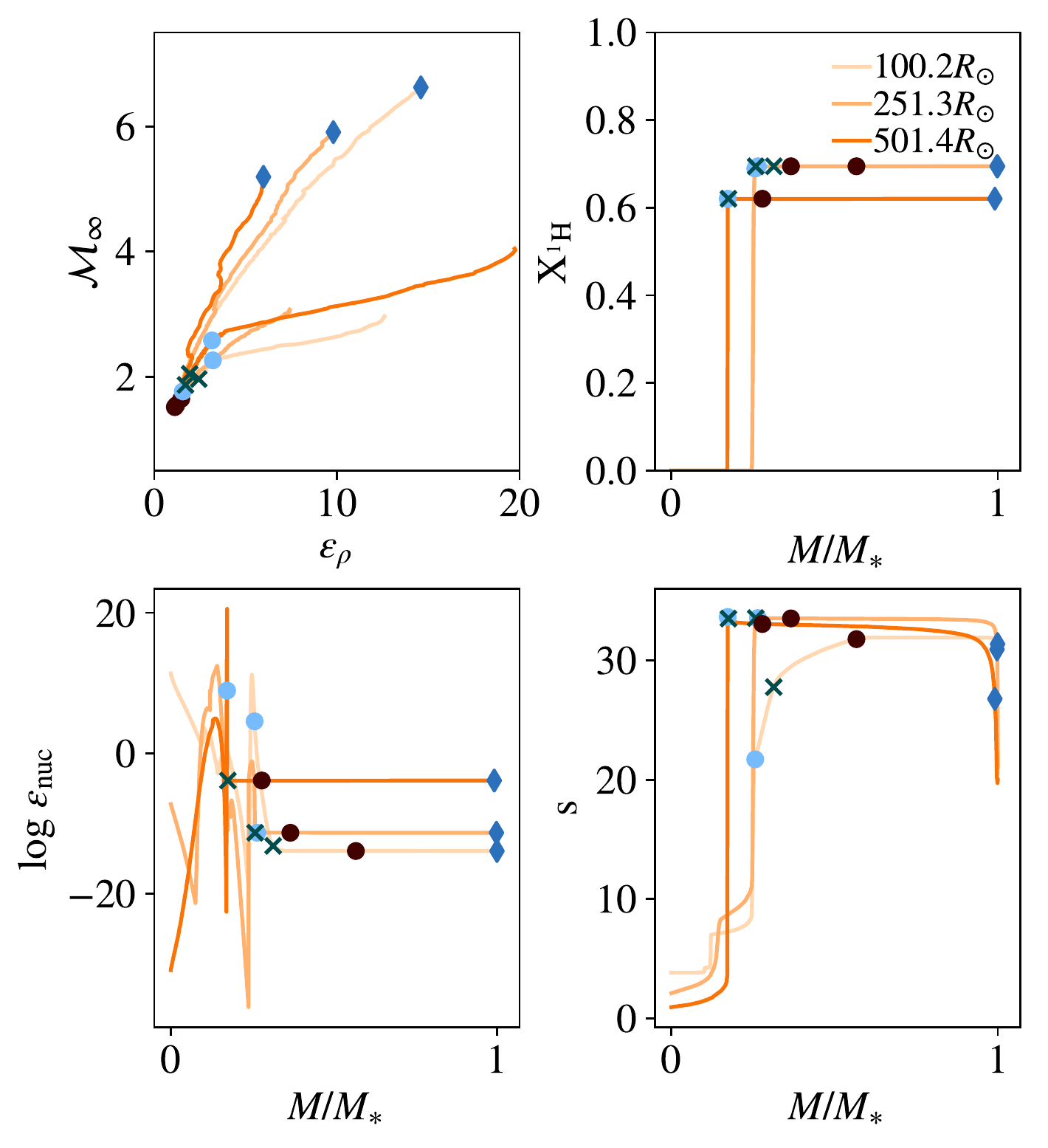}
    \caption{Quantities shown are drawn from stellar models of initial mass $6 M_\odot$ at extents of $100.2 R_\odot$, $251.3 R_\odot$, and $501.4 R_\odot$. For each stellar profile, blue diamonds mark the beginning of inspiral calculations near the limb, blue dots mark the location of $X_\mathrm{^1H} = 0.1$, brown dark dots mark the location of the dynamical boundary, and crosses mark the location at which the $\alpha=1$ criterion for envelope ejection is satisfied. \textit{Upper left:} Raw calculations of characteristic curves for dynamical inspiral in $\mathcal{M}_\infty - \epsilon_\rho$ parameter space with $q_\mathrm{B}=0.3$. Lines descending from the blue diamonds represent dynamical inspiral through envelope material, while shallow tails crossing left to right beyond the inflection point are the same calculations across and beyond the core boundary. Remaining panels reflect various structural quantities used in the literature to discern the core boundary. \textit{Upper right:} Hydrogen mass fraction versus mass. Note that the curve for the $100.2 R_\odot$ profile lies beneath that of the $251.3 R_\odot$. \textit{Lower left:} Nuclear energy generation versus mass. \textit{Lower right:} Entropy versus mass.} \label{fig:hook}
\end{figure}

In Figure \ref{fig:hook}, we use a $6 M_\odot$ primary extended to $\sim100$, $200$, and $500 R_\odot$ to plot raw characteristic curve calculations (upper left), $^1$H mass fraction (upper right), nuclear energy generation (lower left), and entropy (lower right). In the upper left panel, the steeper curves represent the dynamical inspiral phase with calculations beginning at the blue diamonds. These descend from the top right to bottom left, then have a sharp inflection point at or near the minimum Mach value: this is the dynamical boundary, marked by dark brown dots. The tails that then pass from left to right fall outside the applicable range of the drag formalism. The sharp increase in $\epsilon_\rho$ mirrors the steep density gradient which occurs at the core boundary, but doesn't coincide with the location at which the traditional $X_\mathrm{^1H} = 0.1$ criterion is met, marked by blue dots. In all panels, it can be seen that the dynamical boundary precedes the structural core boundary in all cases - this is due to the $R_\mathrm{a}$ dependence in calculating $\epsilon_\rho$, which incorporates the core boundary into the characteristic curve ``before'' the secondary arrives at the core.

For CE events, it is often more desirable to identify the so-called bifurcation point: the location that marks the extent of the remaining material, which may include the core and some envelope remnants, if CE ejection is successful. Estimates for this location are readily calculated using the well-known energy formalism \citep{vandenHeuvel1976, Webbink1984, Livio1988, deKool1990, Iben1993}:
\begin{equation}
    E_\mathrm{bind}(r) = \alpha \Delta E_\mathrm{orb},
\end{equation}
in which $E_\mathrm{bind}(r)$ is the gravitational binding energy of the envelope at $r$, $\Delta E_\mathrm{orb}$ is the change in orbital energy of the secondary from the separation at onset to $r$, and $\alpha$ is an efficiency term of order unity. For cases in which $\alpha=1$, all orbital energy that is lost through inspiral is used to eject the envelope (assuming no additions or losses from other physical processes), and the location $r$ at which they are equated (marked in Figure \ref{fig:hook} by crosses), meaning there has been enough energy deposited to eject the envelope from that point outward, is a loose proxy for the bifurcation point.

The dynamical boundary is not the bifurcation point. Rather, because the drag formalism applies strictly to dynamical inspiral, the dynamical boundary represents the innermost location at which a dynamical inspiral is possible, not accounting for the timescales of energy transport. Comparisons by \citet{Chamandy2019a} show a break in agreement between the drag force as calculated using the drag formalism and that measured in a 3D global hydrodynamic simulation; this break occurs not so much due to changes in local mass ratio, as they suggest, but because the dynamical boundary has been reached and the secondary is entering a self-regulating inspiral, in which the drag formalism is not applicable. 

In general, a secondary that has reached the dynamical boundary has the following possible outcomes: the secondary is plunging in and will merge with the core of the primary, or the secondary is transitioning to a self-regulated inspiral and, if energy considerations permit ejection of the envelope, binarity will be preserved. As such, it is consistent that the dynamical boundary should lie some small distance outside the core and bifurcation point, as in Figure \ref{fig:hook} the dynamical boundary for each model lies external to the location at which the $\alpha=1$ ejection criterion is satisfied. This allows us to apply the drag formalism to the full extent of the envelope as long as the conditions for dynamical inspiral are met. Future work will explore the relationship of the dynamical boundary with the initiation of self-regulated inspiral.

\subsection{Effects and Consequences of EOS} \label{subsec:EOS}

Though the majority of characteristic curves in this work show few or no features, there are exceptions. In a polytropic envelope, any characteristic curve would be featureless and follow the shape seen in the right panel of Figure \ref{fig:paramspace}. Because we use realistic stellar models in which the envelope does not always behave as an ideal gas, the values of $\Gamma_\mathrm{s}$, $\gamma_1$, and $\gamma_3$ may diverge, creating notable features on the curve. 

\begin{figure*}[tbp]
    \figurenum{11}
    \epsscale{1.1}
    \plotone{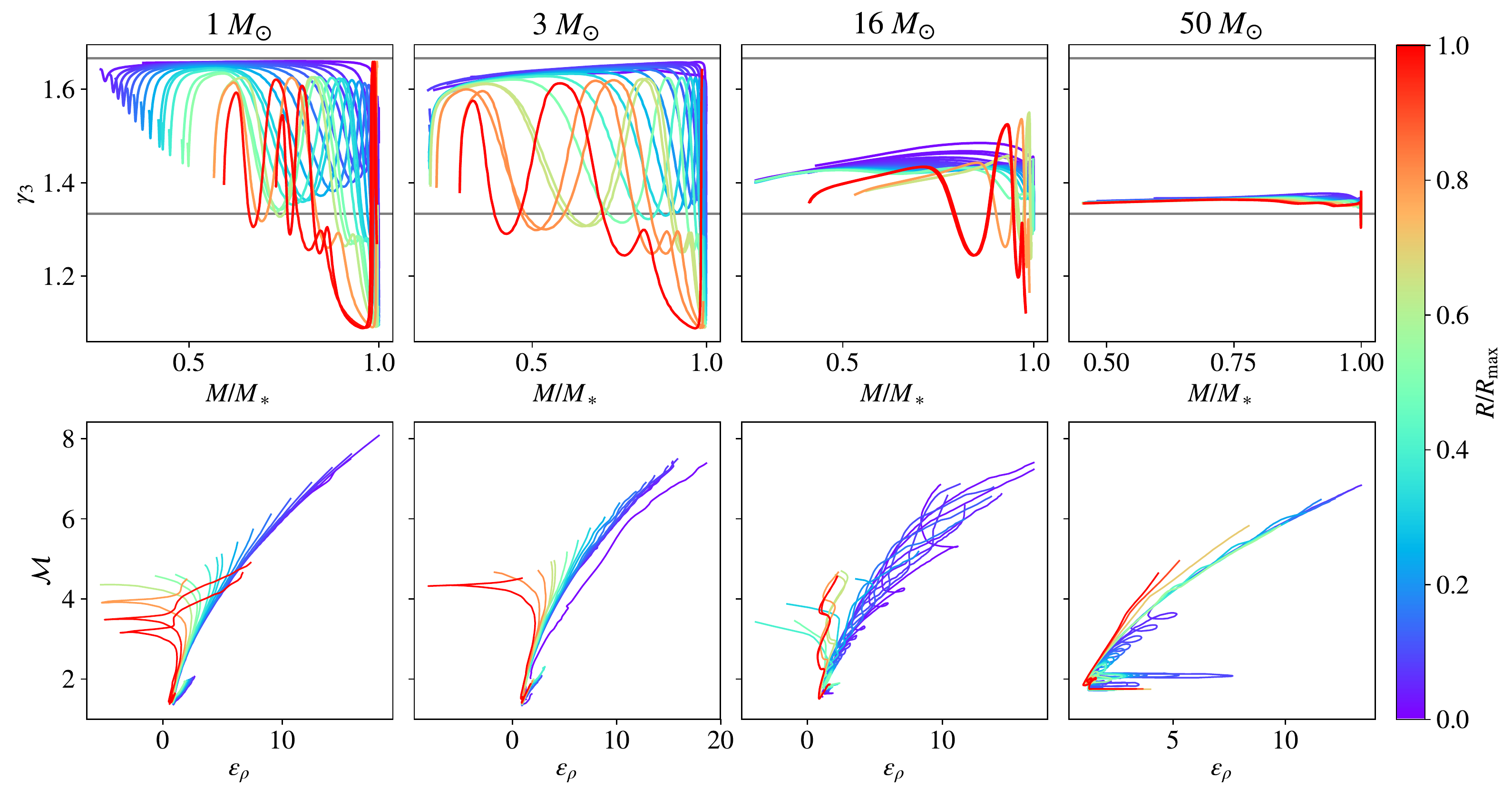}
    \caption{A comparison of $\gamma_3$ values in the envelope and characteristic curves for inspiral with mass ratio 0.2 for stars of initial mass 1, 3, 16, and $50 M_\odot$. Each star is followed from the end of Hydrogen burning to its maximum extent on the giant branch, $R_\mathrm{max}$. Color represents the radius of each stellar profile as a fraction of $R_\mathrm{max}$, with the envelope expanding from purple to red through time. \textit{Upper panels:} Horizontal gray lines are placed at $\gamma_3=5/3$ and $4/3$ for reference. In the lower mass stars, envelopes are seen to evolve from purely convective to bands of convective and radiative regions, with highly compressible regions of partially ionized material in the outer portions of the star near $R_\mathrm{max}$. Large spikes in the outermost regions are density inversions. \textit{Lower panels:} Characteristic curves in $\mathcal{M}_\infty - \epsilon_\rho$ parameter space, matched by color to corresponding EOS curves above. Banded regions do not impact the curves, but density inversions near the limb appear as negative $\epsilon_\rho$ values, precluding these regions from application of the drag formalism. Loops occur in regions where $\Gamma_\mathrm{s}$ diverges from $\gamma_1$.}  \label{fig:eos1}
\end{figure*}

In Figure \ref{fig:eos1}, we compare the $\gamma_3$ values in the envelope against characteristic curves for inspirals of $q_\mathrm{B}=0.2$ in stars of initial mass 1, 3, 16, and $50 M_\odot$ from the end of H-burning (purple) to the maximum radius achieved during the giant branch (red). Demonstrating our baseline, envelopes with $\gamma_3 \sim 5/3$ (purple and dark blue in the four left panels) or $4/3$ (most curves in the rightmost panels) align with the expected featureless morphology of a polytropic curve \citep{2017ApJ...845..173M}. 

In several characteristic curves in the lower panels, loops can be seen, which represent regions in which $\Gamma_\mathrm{s}$ diverges from $\gamma_1$. When such curves are collapsed over the $q$ term as seen above, these loops also collapse. Such variations, as they pertain to drag and accretion coefficients, may be well represented by an averaged featureless curve.

In other profiles, some of the features visible are bands of convective and radiative regions within the same envelope, as well as spikes near the limb that represent density inversions in the outermost envelope. The bands generally do not appear in the characteristic curves, but the density inversions, which are a result of steep temperature gradients in zones of partial ionization \citep{Harpaz1984} that correspond to hydrogen and helium opacity ``bumps" \citep{Sanyal2015,Guzik2018}, fall outside simulated parameters and force $\epsilon_\rho$ values to be negative; thus models that have such density inversions are not appropriate for the drag formalism. It is worth noting that, due to mass loss during onset, the regions containing this feature may possibly be stripped from the star prior to CE, and envelope regions internal to this feature fit comfortably within the established parameter space. However, also worth noting is that there is evidence such density inversions may be a result of 1D simulation that are short-lived (when they appear at all) in 3D simulation and may be non-physical \citep{Jiang2015}. As prescriptions in 1D improve, we may expect an even broader range of models for which the drag formalism is applicable.

\begin{figure*}[tbp]
    \figurenum{12}
    \epsscale{1.1}
    \plotone{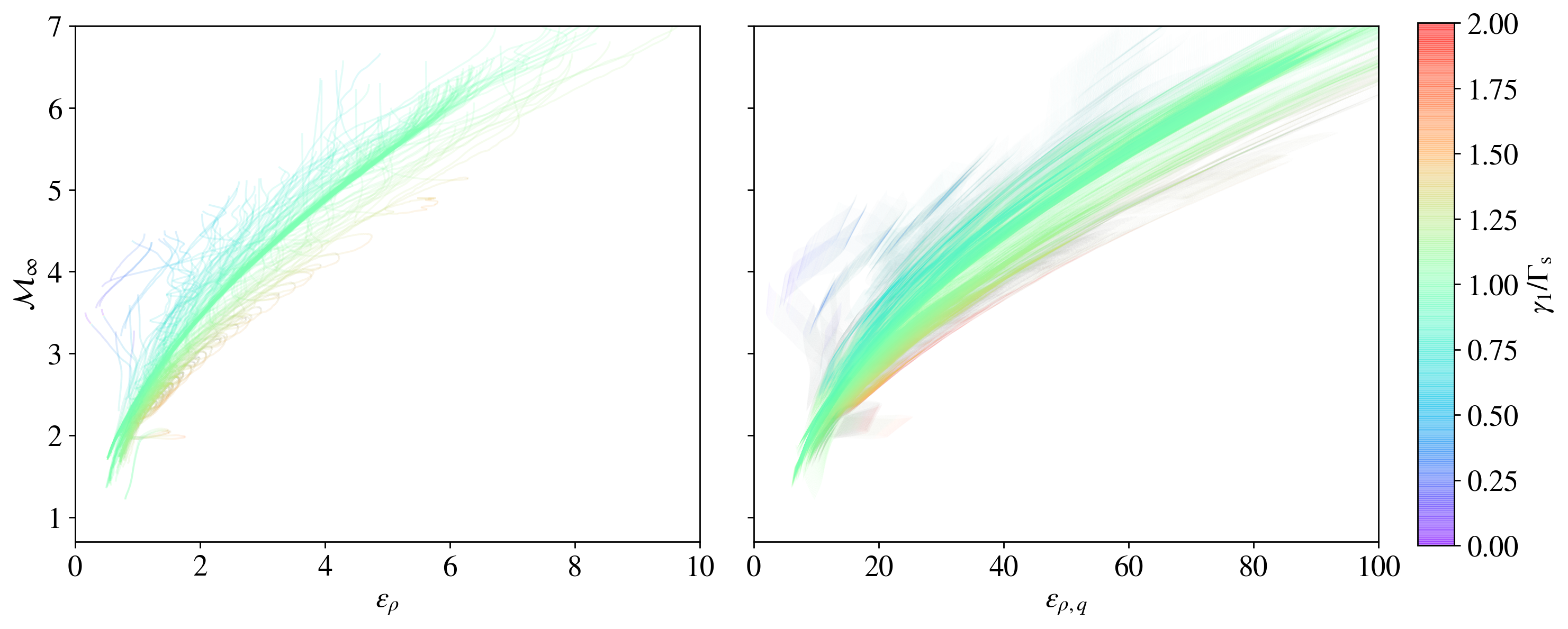}
    \caption{Overlaid values of $\gamma_1 / \Gamma_\mathrm{s}$ for all post-main sequence stellar profiles for masses $1-90 M_\odot$, mapped onto $\mathcal{M}_\infty - \epsilon_\rho$ curves for inspirals with $q_\mathrm{B}=0.1$ (left panel) and collapsed $\mathcal{M}_\infty - \epsilon_{\rho,q}$ curves for $q_\mathrm{B}=0.1-0.35$ (right panel). Intensity of color denotes frequency of incidence of the corresponding ratio. The slope dependence of Equation 10 on $\gamma_1 / \Gamma_\mathrm{s}$ is clearly shown, supporting its use even with realistic stellar profiles. In addition, most characteristic curves throughout the giant branch have $\gamma_1 / \Gamma_\mathrm{s}$ at or near 1, suggesting that a simplified expression like that of Equation 11 may be useful for a general prescriptive framework.}
    \label{fig:gammatracks}
\end{figure*}

In Figure \ref{fig:gammatracks}, we map the ratio $\gamma_1 / \Gamma_\mathrm{s}$ for all post-main sequence stellar profiles in our library using overlaid $\mathcal{M}_\infty - \epsilon_\rho$ tracks calculated with $q_\mathrm{B} = 0.1$ (left panel) and overlaid $\mathcal{M}_\infty - \epsilon_{\rho,q}$ tracks calculated for $q_\mathrm{B} = 0.1-0.35$ (right panel). Increased color saturation indicates increased incidence of the corresponding $\gamma_1 / \Gamma_\mathrm{s}$ value in the tracks. The left panels shows that even with realistic stellar envelopes, for any given mass ratio the slope dependence of Equation 10 on $\gamma_1 / \Gamma_\mathrm{s}$ holds, with higher values to the right and lower values to the left, and that ratios around 1 are most common. In the right panel, we validate this for all $q_\mathrm{B}$ values. This is encouraging, and suggests that a prescriptive parameterization of dynamical inspiral may make use of Equation 11 for simplicity, while covering the most relevant part of parameter space for most cases.

\subsection{Alternative Models: Effects of Metallicity} \label{subsec:z}

In CE events, an analysis of stellar profiles across the axis of metallicity is of interest due to the impact of metallicity on winds, mass loss, and maximum radial extent during late stage stellar evolution. These issues are pronounced in the cases of LIGO binary black hole progenitors due to the need to form massive stellar mass black holes while bringing their giant progenitors into very close proximity. Limitations in the MIST models prevent analysis of stellar envelopes of stars that are very high mass and low metallicity.

To address this in part, we generated models using MESA v10398 \citep{Paxton2011, Paxton2013, Paxton2015, Paxton2017} from $16-90 M_\odot$ with $1/50 Z_\odot$ to apply the same analysis. The self-similarity seen in Section \ref{sec:mapping} holds across this axis of metallicity in high mass giant stars, tracing the same parameter space covered by the MIST models. Due to the uncertainties of stellar models at these masses and metallicities, they did not form the basis of this work, but nonetheless present an encouraging possible application in attempting to model the formation of LIGO-type systems.

\section{Discussion and Conclusions}\label{sec:discussion}

The overall self-similarity shown here in the characteristic curves of dynamical inspiral, and the relatively few limitations and exceptions to that self-similarity, suggest that there may be broad prescriptive applications of these results. One key area where a detailed, prescriptive treatment of common envelope may be useful is in population synthesis studies, which currently depend on variations of the energy formalism to discern the success or failure of envelope ejection. 

Despite its many variations, the energy formalism fails to account for the dynamics of CE in a satisfactory way. CE events generally include several distinct stages:
\begin{itemize}
    \item Onset, which occurs after an initial destabilization of the binary and likely results in some mass loss and desynchronization of the secondary and the envelope;
    \item Dynamical inspiral, in which the secondary plunges quickly, deep into the envelope; 
\end{itemize}
after which a system will merge unless energy conditions for envelope ejection are met, in which case we include:
\begin{itemize}
    \item Self-regulated inspiral, in which the secondary slowly loses orbital energy on a timescale similar to that of the remaining envelope's thermal timescale;
    \item Envelope ejection, in which the outer envelope escapes and the remaining envelope contracts, preserving binarity at some final separation.
\end{itemize}
The outcome of a CE event may be impacted by these stages, beyond what may be accounted for by energy considerations alone. The energy formalism cannot address how the envelope is unbound \citep[see, e.g.][]{Soker1992, Soker2017, Clayton2017, Glanz2018}, as it does not address energy transport unless it is assumed to be instantaneous. In addition, the current energy formalism assumes a change in orbital energy based on the energetics of circular orbits, while the recent work of \citet{Wu2020} suggests that the energetics of a steep spiral plunge may differ significantly.

This work, when combined with the corresponding drag and accretion coefficients from \citet{MacLeod2015a, MacLeod2017b, De2019}, provides the basis for a framework for calculating inspiral trajectories with only basic information about a given binary: the masses of the objects, their separation at CE onset, and the core mass of the primary. This can provide timescales for the duration of dynamical inspiral for a variety of CE events, especially as the dynamical boundary provides a natural end point to dynamical inspiral, but cannot speak to onset or final outcome. To improve such trajectories and make predictions about post-CE outcome, complementary frameworks for mass loss during onset and evolution of self-regulated inspiral are needed, as well as adjustments to the energy formalism that take into account the rate of energy transport within the envelope, such as that done by \citet{Wilson2019} for the low mass regime, and the energetics of non-circular inspiral. Future work will discern if these additions may also be applied in general, without the need for stellar profiles. 

The main conclusions of this work are the following:

\begin{enumerate}
    \item Properties of dynamical inspiral through a broad range of realistic giant branch stellar envelopes are well-described by the dimensionless parameters of the drag formalism (left panel of Figure \ref{fig:paramspace} and Section \ref{sec:mapping}). This allows for the broad application of corresponding drag and accretion coefficients to calculate quantities of interest for dynamical inspirals using basic stellar profiles, rather than requiring hydrodynamic simulations (Subsection \ref{subsec:simresults}).
    \item Characteristic curves of dynamical inspiral in the $\mathcal{M}_\infty - \epsilon_\rho$ parameter space are self-similar across the axes of primary mass, separation (age/radius of primary), and binary mass ratio (Figures \ref{fig:fanpanel}, \ref{fig:qnormfan}, and \ref{fig:paramspace}). Additional work suggests the same holds across metallicity as well (Subsection \ref{subsec:z}). This presents the possibility of a general prescriptive framework that may be applied without the use of a stellar profile, with the addition of treatments for onset, self-regulated inspiral, and energy deposition.
    \item The drag formalism presents a natural termination point for dynamical inspiral: the dynamical boundary, which may be intrinsic to the end of CE via the transition to self-regulated inspiral (Subsection \ref{subsec:hook}). Further work will clarify this relationship.
\end{enumerate}

\acknowledgments

We thank J. Schwab, I. Mandel, J. Andrews, S. Wu, S. Toonen, S. Schr\o der, and A. Murguia-Berthier for helpful discussions. R.W.E. is supported by the Eugene V. Cota-Robles Fellowship, the National Science Foundation (NSF) Graduate Research Fellowship Program (Award \#1339067), and the Heising-Simons Foundation. Any opinions, findings, and conclusions or recommendations expressed in this material are those of the authors and do not necessarily reflect the views of the NSF. R.W.E. and P.M. gratefully acknowledge the support of the Danish National Research Foundation (DNRF132) Niels Bohr Professorship of E.R.-R and the Vera Rubin Presidential Chair for Diversity at UCSC. The authors acknowledge the Kavli Foundation, DNRF, and Dark Cosmology Centre for supporting the 2017 Kavli Summer Program in Astrophysics at the Niels Bohr Institute, which hosted this research in part. R.W.E. also thanks her constant companion M.C. Everson, who was born during the completion of this work, and whose strong opinions shaped and motivated its execution.

\vspace{5mm}
\software{Python,
	MESA \citep{Paxton2011, Paxton2013, Paxton2015, Paxton2017}, 
	matplotlib \citep{Hunter2007}, 
	NumPy \citep{vanderwalt2011}, 
	py\_mesa\_reader \citep{WolfSchwab2017}}

\bibliographystyle{aasjournal}
\bibliography{CEbib}

\begin{thebibliography}{}
\expandafter\ifx\csname natexlab\endcsname\relax\def\natexlab#1{#1}\fi
\providecommand{\url}[1]{\href{#1}{#1}}

\bibitem[{Abbott {et~al.}(2019)Abbott, Abbott, Abbott, Abraham, Acernese,
  Ackley, Adams, Adhikari, Adya, Affeldt, Agathos, Agatsuma, Aggarwal, Aguiar,
  Aiello, Ain, Ajith, Allen, Allocca, Aloy, Altin, Amato, Ananyeva, Anderson,
  Anderson, Angelova, Antier, Appert, Arai, Araya, Areeda, Ar\`ene, Arnaud,
  Arun, Ascenzi, Ashton, Aston, Astone, Aubin, Aufmuth, AultONeal, Austin,
  Avendano, Avila-Alvarez, Babak, Bacon, Badaracco, Bader, Bae, Baker,
  Baldaccini, Ballardin, Ballmer, Banagiri, Barayoga, Barclay, Barish, Barker,
  Barkett, Barnum, Barone, Barr, Barsotti, Barsuglia, Barta, Bartlett, Bartos,
  Bassiri, Basti, Bawaj, Bayley, Bazzan, B\'ecsy, Bejger, Belahcene, Bell,
  Beniwal, Berger, Bergmann, Bernuzzi, Bero, Berry, Bersanetti, Bertolini,
  Betzwieser, Bhandare, Bidler, Bilenko, Bilgili, Billingsley, Birch, Birney,
  Birnholtz, Biscans, Biscoveanu, Bisht, Bitossi, Bizouard, Blackburn,
  Blackman, Blair, Blair, Blair, Bloemen, Bode, Boer, Boetzel, Bogaert, Bondu,
  Bonilla, Bonnand, Booker, Boom, Booth, Bork, Boschi, Bose, Bossie, Bossilkov,
  Bosveld, Bouffanais, Bozzi, Bradaschia, Brady, Bramley, Branchesi, Brau,
  Briant, Briggs, Brighenti, Brillet, Brinkmann, Brisson, Brockill, Brooks,
  Brown, Brunett, Buikema, Bulik, Bulten, Buonanno, Buskulic,
  Bustamante~Rosell, Buy, Byer, Cabero, Cadonati, Cagnoli, Cahillane,
  Calder\'on~Bustillo, Callister, Calloni, Camp, Campbell, Canepa, Cannon, Cao,
  Cao, Capocasa, Carbognani, Caride, Carney, Carullo, Casanueva~Diaz,
  Casentini, Caudill, Cavagli\`a, Cavalier, Cavalieri, Cella, Cerd\'a-Dur\'an,
  Cerretani, Cesarini, Chaibi, Chakravarti, Chamberlin, Chan, Chao, Charlton,
  Chase, Chassande-Mottin, Chatterjee, Chaturvedi, Chatziioannou, Cheeseboro,
  Chen, Chen, Chen, Cheng, Cheong, Chia, Chincarini, Chiummo, Cho, Cho, Cho,
  Christensen, Chu, Chua, Chung, Chung, Ciani, Ciobanu, Ciolfi, Cipriano,
  Cirone, Clara, Clark, Clearwater, Cleva, Cocchieri, Coccia, Cohadon, Cohen,
  Colgan, Colleoni, Collette, Collins, Cominsky, Constancio, Conti, Cooper,
  Corban, Corbitt, Cordero-Carri\'on, Corley, Cornish, Corsi, Cortese, Costa,
  Cotesta, Coughlin, Coughlin, Coulon, Countryman, Couvares, Covas, Cowan,
  Coward, Cowart, Coyne, Coyne, Creighton, Creighton, Cripe, Croquette,
  Crowder, Cullen, Cumming, Cunningham, Cuoco, Canton, D\'alya, Danilishin,
  D'Antonio, Danzmann, Dasgupta, Da~Silva~Costa, Datrier, Dattilo, Dave,
  Davier, Davis, Daw, DeBra, Deenadayalan, Degallaix, De~Laurentis,
  Del\'eglise, Del~Pozzo, DeMarchi, Demos, Dent, De~Pietri, Derby, De~Rosa,
  De~Rossi, DeSalvo, de~Varona, Dhurandhar, D\'{\i}az, Dietrich, Di~Fiore,
  Di~Giovanni, Di~Girolamo, Di~Lieto, Ding, Di~Pace, Di~Palma, Di~Renzo,
  Dmitriev, Doctor, Donovan, Dooley, Doravari, Dorrington, Downes, Drago,
  Driggers, Du, Ducoin, Dupej, Dwyer, Easter, Edo, Edwards, Effler, Ehrens,
  Eichholz, Eikenberry, Eisenmann, Eisenstein, Essick, Estelles, Estevez,
  Etienne, Etzel, Evans, Evans, Fafone, Fair, Fairhurst, Fan, Farinon, Farr,
  Farr, Fauchon-Jones, Favata, Fays, Fazio, Fee, Feicht, Fejer, Feng,
  Fernandez-Galiana, Ferrante, Ferreira, Ferreira, Ferrini, Fidecaro, Fiori,
  Fiorucci, Fishbach, Fisher, Fishner, Fitz-Axen, Flaminio, Fletcher, Flynn,
  Fong, Font, Forsyth, Fournier, Frasca, Frasconi, Frei, Freise, Frey, Frey,
  Fritschel, Frolov, Fulda, Fyffe, Gabbard, Gadre, Gaebel, Gair, Gammaitoni,
  Ganija, Gaonkar, Garcia, Garc\'{\i}a-Quir\'os, Garufi, Gateley, Gaudio, Gaur,
  Gayathri, Gemme, Genin, Gennai, George, George, Gergely, Germain, Ghonge,
  Ghosh, Ghosh, Ghosh, Giacomazzo, Giaime, Giardina, Giazotto, Gill, Giordano,
  Glover, Godwin, Goetz, Goetz, Goncharov, Gonz\'alez, Gonzalez~Castro,
  Gopakumar, Gorodetsky, Gossan, Gosselin, Gouaty, Grado, Graef, Granata,
  Grant, Gras, Grassia, Gray, Gray, Greco, Green, Green, Gretarsson, Groot,
  Grote, Grunewald, Gruning, Guidi, Gulati, Guo, Gupta, Gupta, Gustafson,
  Gustafson, Haegel, Halim, Hall, Hall, Hamilton, Hammond, Haney, Hanke, Hanks,
  Hanna, Hannam, Hannuksela, Hanson, Hardwick, Haris, Harms, Harry, Harry,
  Haster, Haughian, Hayes, Healy, Heidmann, Heintze, Heitmann, Hello, Hemming,
  Hendry, Heng, Hennig, Heptonstall, Hernandez~Vivanco, Heurs, Hild, Hinderer,
  Hoak, Hochheim, Hofman, Holgado, Holland, Holt, Holz, Hopkins, Horst, Hough,
  Howell, Hoy, Hreibi, Huang, Huerta, Huet, Hughey, Hulko, Husa, Huttner,
  Huynh-Dinh, Idzkowski, Iess, Ingram, Inta, Intini, Irwin, Isa, Isac, Isi,
  Iyer, Izumi, Jacqmin, Jadhav, Jani, Janthalur, Jaranowski, Jenkins, Jiang,
  Johnson, Johnson-McDaniel, Jones, Jones, Jones, Jonker, Ju, Junker,
  Kalaghatgi, Kalogera, Kamai, Kandhasamy, Kang, Kanner, Kapadia, Karki,
  Karvinen, Kashyap, Kasprzack, Katsanevas, Katsavounidis, Katzman, Kaufer,
  Kawabe, Keerthana, K\'ef\'elian, Keitel, Kennedy, Key, Khalili, Khan, Khan,
  Khan, Khan, Khazanov, Khursheed, Kijbunchoo, Kim, Kim, Kim, Kim, Kim, Kim,
  Kimball, King, King, Kinley-Hanlon, Kirchhoff, Kissel, Kleybolte, Klika,
  Klimenko, Knowles, Koch, Koehlenbeck, Koekoek, Koley, Kondrashov, Kontos,
  Koper, Korobko, Korth, Kowalska, Kozak, Kringel, Krishnendu, Kr\'olak, Kuehn,
  Kumar, Kumar, Kumar, Kumar, Kuo, Kutynia, Kwang, Lackey, Lai, Lam, Landry,
  Lane, Lang, Lange, Lantz, Lanza, Lartaux-Vollard, Lasky, Laxen, Lazzarini,
  Lazzaro, Leaci, Leavey, Lecoeuche, Lee, Lee, Lee, Lee, Lee, Lee, Lehmann,
  Lenon, Leroy, Letendre, Levin, Li, Li, Li, Li, Lin, Linde, Linker,
  Littenberg, Liu, Liu, Lo, Lockerbie, London, Longo, Lorenzini, Loriette,
  Lormand, Losurdo, Lough, Lousto, Lovelace, Lower, L\"uck, Lumaca, Lundgren,
  Lynch, Ma, Macas, Macfoy, MacInnis, Macleod, Macquet, Maga\~na Sandoval,
  Maga\~na Zertuche, Magee, Majorana, Maksimovic, Malik, Man, Mandic, Mangano,
  Mansell, Manske, Mantovani, Marchesoni, Marion, M\'arka, M\'arka, Markakis,
  Markosyan, Markowitz, Maros, Marquina, Marsat, Martelli, Martin, Martin,
  Martynov, Mason, Massera, Masserot, Massinger, Masso-Reid, Mastrogiovanni,
  Matas, Matichard, Matone, Mavalvala, Mazumder, McCann, McCarthy, McClelland,
  McCormick, McCuller, McGuire, McIver, McManus, McRae, McWilliams, Meacher,
  Meadors, Mehmet, Mehta, Meidam, Melatos, Mendell, Mercer, Mereni, Merilh,
  Merzougui, Meshkov, Messenger, Messick, Metzdorff, Meyers, Miao, Michel,
  Middleton, Mikhailov, Milano, Miller, Miller, Millhouse, Mills,
  Milovich-Goff, Minazzoli, Minenkov, Mishkin, Mishra, Mistry, Mitra,
  Mitrofanov, Mitselmakher, Mittleman, Mo, Moffa, Mogushi, Mohapatra, Montani,
  Moore, Moraru, Moreno, Morisaki, Mours, Mow-Lowry, Mukherjee, Mukherjee,
  Mukherjee, Mukund, Mullavey, Munch, Mu\~niz, Muratore, Murray, Nagar,
  Nardecchia, Naticchioni, Nayak, Neilson, Nelemans, Nelson, Nery, Neunzert,
  Ng, Ng, Nguyen, Nichols, Nielsen, Nissanke, Nitz, Nocera, North, Nuttall,
  Obergaulinger, Oberling, O'Brien, O'Dea, Ogin, Oh, Oh, Ohme, Ohta, Okada,
  Oliver, Oppermann, Oram, O'Reilly, Ormiston, Ortega, O'Shaughnessy, Ossokine,
  Ottaway, Overmier, Owen, Pace, Pagano, Page, Pai, Pai, Palamos, Palashov,
  Palomba, Pal-Singh, Pan, Pang, Pang, Pankow, Pannarale, Pant, Paoletti,
  Paoli, Papa, Parida, Parker, Pascucci, Pasqualetti, Passaquieti, Passuello,
  Patil, Patricelli, Pearlstone, Pedersen, Pedraza, Pedurand, Pele, Penn,
  Perego, Perez, Perreca, Pfeiffer, Phelps, Phukon, Piccinni, Pichot,
  Piergiovanni, Pillant, Pinard, Pirello, Pitkin, Poggiani, Pong, Ponrathnam,
  Popolizio, Porter, Powell, Prajapati, Prasad, Prasai, Prasanna, Pratten,
  Prestegard, Privitera, Prodi, Prokhorov, Puncken, Punturo, Puppo, P\"urrer,
  Qi, Quetschke, Quinonez, Quintero, Quitzow-James, Raab, Radkins, Radulescu,
  Raffai, Raja, Rajan, Rajbhandari, Rakhmanov, Ramirez, Ramos-Buades, Rana,
  Rao, Rapagnani, Raymond, Razzano, Read, Regimbau, Rei, Reid, Reitze, Ren,
  Ricci, Richardson, Richardson, Ricker, Riemenschneider, Riles, Rizzo,
  Robertson, Robie, Robinet, Rocchi, Rolland, Rollins, Roma, Romanelli, Romano,
  Romel, Romie, Rose, Rosi\ifmmode~\acute{n}\else \'{n}\fi{}ska, Rosofsky,
  Ross, Rowan, R\"udiger, Ruggi, Rutins, Ryan, Sachdev, Sadecki, Sakellariadou,
  Salafia, Salconi, Saleem, Salemi, Samajdar, Sammut, Sanchez, Sanchez,
  Sanchis-Gual, Sandberg, Sanders, Santiago, Sarin, Sassolas, Sathyaprakash,
  Saulson, Sauter, Savage, Schale, Scheel, Scheuer, Schmidt, Schnabel,
  Schofield, Sch\"onbeck, Schreiber, Schulte, Schutz, Schwalbe, Scott, Scott,
  Seidel, Sellers, Sengupta, Sennett, Sentenac, Sequino, Sergeev, Setyawati,
  Shaddock, Shaffer, Shahriar, Shaner, Shao, Sharma, Shawhan, Shen, Shink,
  Shoemaker, Shoemaker, ShyamSundar, Siellez, Sieniawska, Sigg, Silva, Singer,
  Singh, Singhal, Sintes, Sitmukhambetov, Skliris, Slagmolen, Slaven-Blair,
  Smith, Smith, Somala, Son, Sorazu, Sorrentino, Souradeep, Sowell, Spencer,
  Srivastava, Srivastava, Staats, Stachie, Standke, Steer, Steinke,
  Steinlechner, Steinlechner, Steinmeyer, Stevenson, Stocks, Stone, Stops,
  Strain, Stratta, Strigin, Strunk, Sturani, Stuver, Sudhir, Summerscales, Sun,
  Sunil, Suresh, Sutton, Swinkels, Szczepa\ifmmode~\acute{n}\else
  \'{n}\fi{}czyk, Tacca, Tait, Talbot, Talukder, Tanner, T\'apai, Taracchini,
  Tasson, Taylor, Thies, Thomas, Thomas, Thondapu, Thorne, Thrane, Tiwari,
  Tiwari, Tiwari, Toland, Tonelli, Tornasi, Torres-Forn\'e, Torrie, T\"oyr\"a,
  Travasso, Traylor, Tringali, Trovato, Trozzo, Trudeau, Tsang, Tse, Tso,
  Tsukada, Tsuna, Tuyenbayev, Ueno, Ugolini, Unnikrishnan, Urban, Usman,
  Vahlbruch, Vajente, Valdes, van Bakel, van Beuzekom, van~den Brand, Van
  Den~Broeck, Vander-Hyde, van Heijningen, van~der Schaaf, van Veggel, Vardaro,
  Varma, Vass, Vas\'uth, Vecchio, Vedovato, Veitch, Veitch, Venkateswara,
  Venugopalan, Verkindt, Vetrano, Vicer\'e, Viets, Vine, Vinet, Vitale, Vo,
  Vocca, Vorvick, Vyatchanin, Wade, Wade, Wade, Walet, Walker, Wallace, Walsh,
  Wang, Wang, Wang, Wang, Wang, Ward, Warden, Warner, Was, Watchi, Weaver, Wei,
  Weinert, Weinstein, Weiss, Wellmann, Wen, Wessel, We\ss{}els, Westhouse,
  Wette, Whelan, White, Whiting, Whittle, Wilken, Williams, Williamson, Willis,
  Willke, Wimmer, Winkler, Wipf, Wittel, Woan, Woehler, Wofford, Worden,
  Wright, Wu, Wysocki, Xiao, Yamamoto, Yancey, Yang, Yap, Yazback, Yeeles, Yu,
  Yu, Yuen, Yvert, Zadro\ifmmode~\dot{z}\else \.{z}\fi{}ny, Zanolin, Zappa,
  Zelenova, Zendri, Zevin, Zhang, Zhang, Zhang, Zhao, Zhou, Zhou, Zhu,
  Zimmerman, Zlochower, Zucker, \& Zweizig}]{Abbott2019}
Abbott, B.~P., Abbott, R., Abbott, T.~D., {et~al.} 2019, Phys. Rev. X, 9,
  031040

\bibitem[{{Bethe} \& {Brown}(1998)}]{Bethe1998}
{Bethe}, H.~A., \& {Brown}, G.~E. 1998, \apj, 506, 780

\bibitem[{Bondi \& Hoyle(1944)}]{Bondi1944}
Bondi, H., \& Hoyle, F. 1944, MNRAS, 104, 273

\bibitem[{Chamandy {et~al.}(2019{\natexlab{a}})Chamandy, Blackman, Frank,
  Carroll-Nellenback, Zou, \& Tu}]{Chamandy2019a}
Chamandy, L., Blackman, E.~G., Frank, A., {et~al.} 2019{\natexlab{a}}, MNRAS,
  490, 3727

\bibitem[{Chamandy {et~al.}(2019{\natexlab{b}})Chamandy, Tu, Blackman,
  Carroll-Nellenback, Frank, Liu, \& Nordhaus}]{Chamandy2019b}
Chamandy, L., Tu, Y., Blackman, E.~G., {et~al.} 2019{\natexlab{b}}, MNRAS, 486,
  1070

\bibitem[{Chamandy {et~al.}(2018)Chamandy, Frank, Blackman, Carroll-Nellenback,
  Liu, Tu, Nordhaus, Chen, \& Peng}]{Chamandy2018}
Chamandy, L., Frank, A., Blackman, E.~G., {et~al.} 2018, MNRAS, 480, 1898

\bibitem[{Choi {et~al.}(2016)Choi, Dotter, Conroy, Cantiello, Paxton, \&
  Johnson}]{Choi2016}
Choi, J., Dotter, A., Conroy, C., {et~al.} 2016, ApJ, 823, 102

\bibitem[{Clayton {et~al.}(2017)Clayton, Podsiadlowski, Ivanova, \&
  Justham}]{Clayton2017}
Clayton, M., Podsiadlowski, P., Ivanova, N., \& Justham, S. 2017, MNRAS, 470,
  1788

\bibitem[{{Dan} {et~al.}(2011){Dan}, {Rosswog}, {Guillochon}, \&
  {Ramirez-Ruiz}}]{2011ApJ...737...89D}
{Dan}, M., {Rosswog}, S., {Guillochon}, J., \& {Ramirez-Ruiz}, E. 2011, \apj,
  737, 89

\bibitem[{De {et~al.}(2020)De, MacLeod, Everson, Mandel, \&
  Ramirez-Ruiz}]{De2019}
De, S., MacLeod, M., Everson, R.~W., Mandel, I., \& Ramirez-Ruiz, E. 2020, ApJ,
  897, 130

\bibitem[{{de Kool}(1990)}]{deKool1990}
{de Kool}, M. 1990, ApJ, 358, 189

\bibitem[{Dotter(2016)}]{Dotter2016}
Dotter, A. 2016, ApJS, 222, 8

\bibitem[{Fragos {et~al.}(2019)Fragos, Andrews, Ramirez-Ruiz, Meynet, Kalogera,
  Taam, \& Zezas}]{Fragos2019}
Fragos, T., Andrews, J.~J., Ramirez-Ruiz, E., {et~al.} 2019, ApJ, 883, L45

\bibitem[{Glanz \& Perets(2018)}]{Glanz2018}
Glanz, H., \& Perets, H.~B. 2018, MNRAS, 478, L12

\bibitem[{Grichener {et~al.}(2018)Grichener, Sabach, \& Soker}]{Grichener2018}
Grichener, A., Sabach, E., \& Soker, N. 2018, MNRAS, 478, 1818

\bibitem[{Guzik {et~al.}(2018)Guzik, Fontes, \& Fryer}]{Guzik2018}
Guzik, J., Fontes, C., \& Fryer, C. 2018, Atoms, 6, 31

\bibitem[{Harpaz(1984)}]{Harpaz1984}
Harpaz, A. 1984, MNRAS, 210, 633

\bibitem[{Hoyle \& Lyttleton(1939)}]{Hoyle1939}
Hoyle, F., \& Lyttleton, R. 1939, PCPS, 35, 405

\bibitem[{Hunter(2007)}]{Hunter2007}
Hunter, J.~D. 2007, Computing In Science \& Engineering, 9, 90

\bibitem[{Iaconi {et~al.}(2017)Iaconi, Reichardt, Staff, De~Marco, Passy,
  Price, Wurster, \& Herwig}]{Iaconi2016}
Iaconi, R., Reichardt, T., Staff, J., {et~al.} 2017, MNRAS, 464, 4028

\bibitem[{Iben \& Livio(1993)}]{Iben1993}
Iben, Icko, J., \& Livio, M. 1993, PASP, 105, 1373

\bibitem[{Ivanova(2016)}]{Ivanova2016a}
Ivanova, N. 2016, in IAU Symp. 329, The Lives and Death-Throes of Massive
  Stars, ed. J.~Eldridge, J.~Bray, L.~McClelland, \& L.~Xiao, Vol.~12, 199--206

\bibitem[{Ivanova \& Nandez(2016)}]{Ivanova2016}
Ivanova, N., \& Nandez, J.~L. 2016, MNRAS, 462, 362

\bibitem[{Ivanova {et~al.}(2013)Ivanova, Justham, Chen, {De Marco}, Fryer,
  Gaburov, Ge, Glebbeek, Han, Li, Lu, Marsh, Podsiadlowski, Potter, Soker,
  Taam, Tauris, van~den Heuvel, \& Webbink}]{Ivanova2013}
Ivanova, N., Justham, S., Chen, X., {et~al.} 2013, A{\&}ARv, 21, 59

\bibitem[{{Jiang} {et~al.}(2015){Jiang}, Cantiello, Bildsten, Quataert, \&
  Blaes}]{Jiang2015}
{Jiang}, Y.~F., Cantiello, M., Bildsten, L., Quataert, E., \& Blaes, O. 2015,
  ApJ, 813, 74

\bibitem[{{Lee} \& {Ramirez-Ruiz}(2007)}]{Lee2007}
{Lee}, W.~H., \& {Ramirez-Ruiz}, E. 2007, New Journal of Physics, 9, 17

\bibitem[{Livio \& Soker(1988)}]{Livio1988}
Livio, M., \& Soker, N. 1988, ApJ, 329, 764

\bibitem[{MacLeod {et~al.}(2017{\natexlab{a}})MacLeod, Antoni, Murgia-Berthier,
  Macias, \& Ramirez-Ruiz}]{MacLeod2017b}
MacLeod, M., Antoni, A., Murgia-Berthier, A., Macias, P., \& Ramirez-Ruiz, E.
  2017{\natexlab{a}}, ApJ, 838, 56

\bibitem[{MacLeod \& Loeb(2020{\natexlab{a}})}]{MacLeod2020a}
MacLeod, M., \& Loeb, A. 2020{\natexlab{a}}, ApJ, 893, 106

\bibitem[{MacLeod \& Loeb(2020{\natexlab{b}})}]{MacLeod2020b}
---. 2020{\natexlab{b}}, ApJ, 895, 29

\bibitem[{MacLeod {et~al.}(2017{\natexlab{b}})MacLeod, Macias, Ramirez-Ruiz,
  Grindlay, Batta, \& Montes}]{MacLeod2017a}
MacLeod, M., Macias, P., Ramirez-Ruiz, E., {et~al.} 2017{\natexlab{b}}, ApJ,
  835, 282

\bibitem[{MacLeod {et~al.}(2018)MacLeod, Ostriker, \& Stone}]{MacLeod2018}
MacLeod, M., Ostriker, E.~C., \& Stone, J.~M. 2018, ApJ, 863, 5

\bibitem[{MacLeod \& Ramirez-Ruiz(2015{\natexlab{a}})}]{MacLeod2015a}
MacLeod, M., \& Ramirez-Ruiz, E. 2015{\natexlab{a}}, ApJ, 803, 41

\bibitem[{MacLeod \& Ramirez-Ruiz(2015{\natexlab{b}})}]{MacLeod2015b}
---. 2015{\natexlab{b}}, ApJ, 798, L19

\bibitem[{Mandel \& {de Mink}(2016)}]{Mandel2016}
Mandel, I., \& {de Mink}, S.~E. 2016, MNRAS, 458, 2634

\bibitem[{{Murguia-Berthier} {et~al.}(2017){Murguia-Berthier}, {MacLeod},
  {Ramirez-Ruiz}, {Antoni}, \& {Macias}}]{2017ApJ...845..173M}
{Murguia-Berthier}, A., {MacLeod}, M., {Ramirez-Ruiz}, E., {Antoni}, A., \&
  {Macias}, P. 2017, \apj, 845, 173

\bibitem[{Nandez \& Ivanova(2016)}]{Nandez2016}
Nandez, J. L.~A., \& Ivanova, N. 2016, MNRAS, 460, 3992

\bibitem[{Nandez {et~al.}(2014)Nandez, Ivanova, \& J.~C.~Lombardi}]{Nandez2014}
Nandez, J. L.~A., Ivanova, N., \& J.~C.~Lombardi, J. 2014, ApJ, 786, 39

\bibitem[{Ohlmann {et~al.}(2017)Ohlmann, {R\"opke}, {Pakmor}, \&
  {Springel}}]{Ohlmann2017}
Ohlmann, S.~T., {R\"opke}, F.~K., {Pakmor}, R., \& {Springel}, V. 2017, A\&A,
  599, A5

\bibitem[{Ohlmann {et~al.}(2016{\natexlab{a}})Ohlmann, Röpke, Pakmor, \&
  Springel}]{Ohlmann2016a}
Ohlmann, S.~T., Röpke, F.~K., Pakmor, R., \& Springel, V. 2016{\natexlab{a}},
  ApJ, 816, L9

\bibitem[{Ohlmann {et~al.}(2016{\natexlab{b}})Ohlmann, Röpke, Pakmor,
  Springel, \& Müller}]{Ohlmann2016b}
Ohlmann, S.~T., Röpke, F.~K., Pakmor, R., Springel, V., \& Müller, E.
  2016{\natexlab{b}}, MNRAS, 462, L121

\bibitem[{Paczynski(1976)}]{Paczynski1976}
Paczynski, B. 1976, in IAU Symp. 73, Structure and Evolution of Close Binary
  Systems, ed. P.~Eggleton, S.~Mitton, \& J.~Whelan (Dordrecht: D. Reidel),
  75--80

\bibitem[{Passy {et~al.}(2012)Passy, Marco, Fryer, Herwig, Diehl, Oishi, Low,
  Bryan, \& Rockefeller}]{Passy2012}
Passy, J.-C., Marco, O.~D., Fryer, C.~L., {et~al.} 2012, ApJ, 744, 52

\bibitem[{Paxton {et~al.}(2011)Paxton, Bildsten, Dotter, Herwig, Lesaffre, \&
  Timmes}]{Paxton2011}
Paxton, B., Bildsten, L., Dotter, A., {et~al.} 2011, ApJS, 192, 3

\bibitem[{Paxton {et~al.}(2013)Paxton, Cantiello, Arras, Bildsten, Brown,
  Dotter, Mankovich, Montgomery, Stello, Timmes, \& Townsend}]{Paxton2013}
Paxton, B., Cantiello, M., Arras, P., {et~al.} 2013, ApJS, 208, 4

\bibitem[{Paxton {et~al.}(2015)Paxton, Marchant, Schwab, Bauer, Bildsten,
  Cantiello, Dessart, Farmer, Hu, Langer, Townsend, Townsley, \&
  Timmes}]{Paxton2015}
Paxton, B., Marchant, P., Schwab, J., {et~al.} 2015, ApJS, 220, 15

\bibitem[{Paxton {et~al.}(2017)Paxton, Schwab, Bauer, Bildsten, Blinnikov,
  Duffell, Farmer, Goldberg, Marchant, Sorokina, Thoul, Townsend, \&
  Timmes}]{Paxton2017}
Paxton, B., Schwab, J., Bauer, E.~B., {et~al.} 2017, ApJS, 234, 34

\bibitem[{{Postnov} \& {Yungelson}(2014)}]{Postnov2014}
{Postnov}, K.~A., \& {Yungelson}, L.~R. 2014, Living Reviews in Relativity, 17,
  3

\bibitem[{Prust \& Chang(2019)}]{Prust2019}
Prust, L.~J., \& Chang, P. 2019, MNRAS, 486, 5809

\bibitem[{Reichardt {et~al.}(2019)Reichardt, {De Marco}, Iaconi, Tout, \&
  Price}]{Reichardt2019}
Reichardt, T.~A., {De Marco}, O., Iaconi, R., Tout, C.~A., \& Price, D.~J.
  2019, MNRAS, 484, 631

\bibitem[{Ricker \& Taam(2008)}]{Ricker2008}
Ricker, P.~M., \& Taam, R.~E. 2008, ApJ, 672, L41

\bibitem[{Ricker \& Taam(2012)}]{Ricker2012}
---. 2012, ApJ, 746, 74

\bibitem[{Rodriguez {et~al.}(2018)Rodriguez, Amaro-Seoane, Chatterjee, \&
  Rasio}]{Rodriguez2018}
Rodriguez, C.~L., Amaro-Seoane, P., Chatterjee, S., \& Rasio, F.~A. 2018, Phys.
  Rev. Lett., 120, 151101

\bibitem[{Samsing(2018)}]{Samsing2018}
Samsing, J. 2018, Phys. Rev. D, 97, 103014

\bibitem[{{Sana} {et~al.}(2012){Sana}, {de Mink}, {de Koter}, {Langer},
  {Evans}, {Gieles}, {Gosset}, {Izzard}, {Le Bouquin}, \&
  {Schneider}}]{Sana2012}
{Sana}, H., {de Mink}, S.~E., {de Koter}, A., {et~al.} 2012, Science, 337, 444

\bibitem[{Sanyal {et~al.}(2015)Sanyal, Grassitelli, Langer, \&
  Bestenlehner}]{Sanyal2015}
Sanyal, D., Grassitelli, L., Langer, N., \& Bestenlehner, J.~M. 2015, A{\&}A,
  580, A20

\bibitem[{Shiber {et~al.}(2019)Shiber, Iaconi, {De Marco}, \&
  Soker}]{Shiber2019}
Shiber, S., Iaconi, R., {De Marco}, O., \& Soker, N. 2019, MNRAS, 488, 5615

\bibitem[{Soker(1992)}]{Soker1992}
Soker, N. 1992, ApJ, 399, 185

\bibitem[{Soker(2017)}]{Soker2017}
---. 2017, MNRAS, 471, 4839

\bibitem[{Staff {et~al.}(2015)Staff, De~Marco, Macdonald, Galaviz, Passy,
  Iaconi, \& Low}]{Staff2015}
Staff, J.~E., De~Marco, O., Macdonald, D., {et~al.} 2015, MNRAS, 455, 3511

\bibitem[{Staff {et~al.}(2016)Staff, De~Marco, Wood, Galaviz, \&
  Passy}]{Staff2016}
Staff, J.~E., De~Marco, O., Wood, P., Galaviz, P., \& Passy, J.-C. 2016, MNRAS,
  458, 832

\bibitem[{Taam \& Ricker(2010)}]{Taam2010}
Taam, R.~E., \& Ricker, P.~M. 2010, New Ast. Rev., 54, 65

\bibitem[{Taam \& Sandquist(2000)}]{Taam2000}
Taam, R.~E., \& Sandquist, E.~L. 2000, ARA{\&}A, 38, 113

\bibitem[{Tauris \& Dewi(2001)}]{Tauris2001}
Tauris, T.~M., \& Dewi, J. D.~M. 2001, A{\&}A, 369, 170

\bibitem[{{Toonen} {et~al.}(2016){Toonen}, {Hamers}, \& {Portegies
  Zwart}}]{Toonen2016}
{Toonen}, S., {Hamers}, A., \& {Portegies Zwart}, S. 2016, Comput. Astrophys.
  \& Cosm., 3, 6

\bibitem[{{van den Heuvel}(1976)}]{vandenHeuvel1976}
{van den Heuvel}, E.~P.~J. 1976, in IAU Symp. 73, Structure and Evolution of
  Close Binary Systems, ed. P.~Eggleton, S.~Mitton, \& J.~Whelan (Dordrecht: D.
  Reidel), 35

\bibitem[{van~der Walt {et~al.}(2011)van~der Walt, Colbert, \&
  Varoquaux}]{vanderwalt2011}
van~der Walt, S., Colbert, S.~C., \& Varoquaux, G. 2011, Computing in Science
  \& Engineering, 13, 22

\bibitem[{{Vigna-G{\'o}mez} {et~al.}(2020){Vigna-G{\'o}mez}, {MacLeod},
  {Neijssel}, {Broekgaarden}, {Justham}, {Howitt}, {de Mink}, \& {Mand
  el}}]{2020arXiv200109829V}
{Vigna-G{\'o}mez}, A., {MacLeod}, M., {Neijssel}, C.~J., {et~al.} 2020, arXiv
  e-prints, arXiv:2001.09829

\bibitem[{Webbink(1984)}]{Webbink1984}
Webbink, R.~F. 1984, ApJ, 277, 355

\bibitem[{Wilson \& Nordhaus(2019)}]{Wilson2019}
Wilson, E.~C., \& Nordhaus, J. 2019, MNRAS, 485, 4492

\bibitem[{Wilson \& Nordhaus(2020)}]{Wilson2020}
---. 2020, arXiv: 2006.09360

\bibitem[{Wolf \& Schwab(2017)}]{WolfSchwab2017}
Wolf, B., \& Schwab, J. 2017, wmwolf/py\textunderscore mesa\textunderscore
  reader: Interact with MESA Output,  Zenodo, doi:10.5281/zenodo.826958

\bibitem[{Wu {et~al.}(2020)Wu, Everson, Schneider, Podsiadlowski, \&
  Ramirez-Ruiz}]{Wu2020}
Wu, S., Everson, R.~W., Schneider, F. R.~N., Podsiadlowski, P., \&
  Ramirez-Ruiz, E. 2020, arXiv: 2006.01940

\end{thebibliography}

\end{document}